\documentclass[a4paper,11pt]{article}

\usepackage{graphicx}
\usepackage{fullpage}
\usepackage{libertine}
\usepackage{color}

\usepackage{caption}
\usepackage{tabularx}
\usepackage{subcaption}
\usepackage{enumerate}
\usepackage{color}
\usepackage{float}
\usepackage{graphicx}

\usepackage{amsmath,amsfonts,amssymb,amsthm}
\usepackage{multirow}
\usepackage[svgnames]{xcolor}
\usepackage[capitalise,nameinlink]{cleveref}

\usepackage{subcaption}

\newtheorem{theorem}{Theorem}[section]
\newtheorem{corollary}[theorem]{Corollary}

\newtheorem{lemma}[theorem]{Lemma}

\theoremstyle{definition}

\title{Diversity-seeking swap games in networks}

\usepackage{authblk}
\author[1]{Yaqiao Li}
\author[2]{Lata Narayanan}
\author[2]{Jaroslav Opatrny}
\author[3]{Yi Tian Xu}
\affil[1]{Faculty of CSCE, Shenzhen University of Advanced Technology, China}
\affil[2]{Department of CSSE, Concordia University, Montreal, Canada}\affil[3]{Shenzhen, China}
\date{}
\newtheorem{observation}{Observation}
\usepackage{mathtools}
  
\DeclarePairedDelimiter\floor{\lfloor}{\rfloor}
\newcommand{\cX}{\mathcal X}    
\newcommand{\cT}{\mathcal T}    
\newcommand{\N}{\mathbb{N}}

\newcommand{\norm}[1]{\left\Vert#1\right\Vert}

\DeclareMathOperator{\OPT}{OPT}
\DeclareMathOperator{\eq}{EQ}   
\DeclareMathOperator{\sw}{SW}   
\DeclareMathOperator{\ce}{CE}     
\DeclareMathOperator{\ev}{EV}     
\DeclareMathOperator{\nv}{NV}    
\DeclareMathOperator{\doi}{DOI}   
\DeclareMathOperator{\doic}{DOIC}   
\DeclareMathOperator{\doit}{DOIT}   
\newcommand{\poa}{PoA}
\newcommand{\pos}{PoS}
\newcommand{\wdoic}{wDoIC}    %
\newcommand{\wdoit}{wDoIT}    %

\newcommand{\bo}{\bf 1}
\newcommand{\bt}{\bf 2}

\newcommand{\bd}{\bf .}
\newcommand{\mr}[1]{\textcolor{red}{#1}}
\newcommand{\mb}[1]{\textcolor{blue}{#1}}
\newcommand{\mg}[1]{\textcolor{green}{#1}}

\newcommand{\geqas}{\gtrapprox}
\newcommand{\leqas}{\lessapprox}
\newcommand{\eqas}{\simeq}

\begin{document}

\maketitle

\begin{abstract}  
  Schelling games use a game-theoretic approach to study the phenomenon of residential segregation as originally modeled by Schelling \cite{schelling1969models,schelling1971dynamic}. Inspired by the recent  increase in the number of people and businesses preferring and promoting diversity, we propose  swap games under three {\em diversity-seeking} utility functions: the binary utility of an agent is 1 if it has a neighbor of a different type, and 0 otherwise; the difference-seeking utility of an agent is equal to the number of its neighbors of a different type; the variety-seeking utility of an agent is equal to the number of types different from its own in its neighborhood. We consider four {\em global} measures of diversity:  degree of integration, number of colorful edges, neighborhood variety, and  evenness, and prove asymptotically tight or almost tight bounds on the price of anarchy with respect to these measures on both general graphs, as well as on cycles, cylinders, and tori that model residential neighborhoods. 
  We complement our theoretical results with  simulations of our swap games starting either from random placements of agents, or from segregated placements. 
  Our simulation results are generally consistent with our theoretical results, showing that segregation is effectively removed when agents are diversity-seeking; however strong diversity, such as measured by neighborhood variety and evenness, is harder to achieve by our swap games. 
\end{abstract}

\section{Introduction}

In his pioneering work, Schelling  \cite{schelling1969models,schelling1971dynamic} showed that even small individual and local preferences for neighbours of the same type can lead to global segregation. Schelling's work has been hugely influential in sociology and economics and his findings were corroborated in a number of subsequent studies.
In recent years, motivated by Schelling's initial model, computer scientists have been studying related problems  
using the tools of algorithmic game theory: strategic agents of different {\em types} are located at the vertices of a graph, and move to new positions to improve their own {\em utility}. Many variations of utility functions that can generally be described as {\em similarity-seeking} or {\em homophilic}
have been studied \cite{ijcai2022p22,bullinger2021welfare,chauhan2018schelling,echzell2019convergence,schelling-journal,kanellopoulos2020modified}.  Two kinds of games have been considered: in a {\em jump} game, an agent moves to a previously unoccupied location to improve its utility, and in a {\em swap} game,  agents of different types swap locations if it would increase both their utilities.  Researchers have studied the computational complexity of finding equilibria and topologies in which a Nash equilibrium always exists, as well as the efficiency at equilibrium using measures such as the price of anarchy and  stability.

Data from the General Social Survey \cite{smith2019general},  conducted in the US since 1950, show that the percentage of people preferring diverse neighbourhoods has been steadily increasing. A large number of studies show that teams composed of people with diverse backgrounds and skill sets lead to better outcomes for business; see for example
\cite{elia2019impact,filbeck2017does,freeman2015collaborating,loyd2013social}. 
Thus, many governments and private enterprises actively promote increased diversity as being beneficial to both societal harmony and efficient functioning of institutions. This recently motivated some researchers to study strategic games with agents that aim to increase the diversity in their neighbourhood \cite{NS23,ijcai2022p12,friedrich2023single}. Understanding what utility functions contribute to greater global diversity could help in defining appropriate incentives for businesses and governments.

In this paper, we  continue the study of   {\em diversity-seeking} utility functions in strategic games. The meaning of diversity is subjective, and  an agent's perception of a diverse local neighbourhood may be based on several factors. We study swap games based on three different and natural  diversity-seeking utility functions.  The first and simplest utility function  $U_b$, called  {\em binary}, is 1 if an agent has at least one neighbour of a different type and 0 otherwise. The second utility function $U_\#$, called  {\em difference-seeking}, is equal to the number of neighbours of a different type. The last utility function $U_\tau$, called {\em variety-seeking}, is equal to the number of neighbouring  types distinct from the agent's type.

Determining whether our swap games result in {\em global diversity}  requires grappling with the definition of the global diversity in a network.  In the literature on Schelling games, the main measure of diversity considered so far is the so-called {\em degree of integration}, that is, the fraction of agents with at least one neighbour of a different type. We view this as a minimal definition of diversity. 
In this paper, we introduce several other measures of diversity. 
We study  the number of {\em colorful edges}, that is, the number of heterophilic connections between agents, and the {\em neighborhood variety}, which is the average number of types of agents in an agent's neighborhood. Finally, we study {\em evenness}, which measures the divergence of an assignment from a uniform 
distribution of neighbours of different types. We note that the degree of integration, colorful edges, and neighborhood variety correspond to the social welfare of the binary, difference-seeking, and variety-seeking utility functions respectively. Finally, we generalize the notion of $DoI$ in two ways:  we measure the  fraction of agents with {\em at least $k$ neighbours of a different type}, as well as  the  fraction of agents that have {\em neighbours of at least $k$ different types}.


\subsection{Our results} \label{sec:results}

It is easy to see that on arbitrary graphs and with arbitrarily asymmetric 
agents, the price of anarchy with respect to our diversity measures  can be unbounded. Thus, we concentrate on swap games  by {\em equitable agents}, i.e., the number of agents of different types differ by at most 1.  We show the following results:

    (1) We show that the swap game is a \emph{potential game} for all graphs under any of the three utility functions $U_b$, $U_\#$, $U_\tau$. Therefore, improving response dynamics in all these games always results in an equilibrium assignment where no pair of agents have the incentive to swap their locations.

    (2)  For all three swap games on general graphs, we show an asymptotically tight bound of $t/(t-1)$ on the price of anarchy (PoA) with respect to the degree of integration,  where $t$ is the number of types of agents, thus showing that segregation is effectively removed by a swap game under all three of our utility functions. For neighborhood variety on general graphs and for evenness on regular graphs, we show an asymptotically  tight bound of $t$ on the the price of anarchy for all three utility functions, showing that these diversity measures are difficult to optimize by our swap games in the worst case. 
    
    For colorful edges, we show a tight bound of $\Delta t/(t-1)$ on the PoA for $U_b$, where $\Delta$ is the maximum degree. We show that the PoA is at most $\Delta$ for $U_\tau$ for all graphs,  and at most $2$ for $U_\#$ for regular graphs. Furthermore, the PoA approaches 1 for both $U_\tau$ and $U_\#$ on regular graphs as the number of types increases.

    (3) For cycles, cylinders, and tori that model residential neighborhoods, we further improve the PoA bounds of the social welfare and colorful edges. 
    For example, the PoA for colorful edges for $U_\tau$ on the torus is shown to be exactly $24/7$, which means that on average an agent in the worst equilibrium of $U_\tau$ has $4/(24/7) \approx 1.17$ colorful edges, instead of 4, as implied by the general upper bound mentioned above.     
    Our study also shows that the worst equilibria are often determined by interesting \emph{tilings}. 

    (4) We run simulations of swap games with our three utility functions on tori, with random initial placements as well as with segregated initial placements. 
    Overall, our simulation results are consistent with our theoretical results, showing that segregation can be effectively removed when agents become diversity-seeking, and that high degree of integration and colorful edges are easier to achieve than high neighborhood variety and evenness.

\subsection{Related work} \label{sec:related}

In the last decade the random process introduced by Schelling has been theoretically analyzed  in the computer science literature, e.g.,  
\cite{brandt2012analysis,STOC2012Circleanalysis}. The expected size of the resulting segregated neighbourhoods was shown to be polynomial in the size of the neighbourhood on the line \cite{brandt2012analysis} and exponential in its size on the grid \cite{STOC2012Circleanalysis}. 
Zhang \cite{zhang2004dynamic} proposed to study the changes in a neighbourhood as a strategic game played by two types of agents  rather than as a random process. Different utility functions for the agents were studied in \cite{chauhan2018schelling}, and generalizations to $k$ types of agents were studied in \cite{echzell2019convergence}.
Agarwal et al. \cite{schelling-journal} studied jump games  for $k$ types of agents in which agents always seek to increase the fraction of agents of their own type. They modelled location preferences by considering two 
classes of agents: those in the first class  are strategic and aim to maximize their utility, while those in the second class are stubborn, not 
changing their initial location.
The utility function considered by Agarwal et al. \cite{agarwal2020swap} for swap games is similar to that of Elkind et al. \cite{schelling-journal}. 
A new measure of global diversity, called the  {\em degree of integration}, was introduced there 
and  results on the price of anarchy and stability with respect to this measure were given. 
Kreisel et al. \cite{kreisel2021equilibria} showed that determining the existence of equilibria is NP hard in jump and swap games even if all the agents are strategic. The influence of the underlying topology and locality in swap games was studied in Bil{\`o} et al. \cite{bilo2022topological}.

Diversity hedonic games were studied in \cite{bredereck2019hedonic} in which some agents have homophilic preferences, while others have heterophilic preferences. However, hedonic games are different to our model as in hedonic games agents form pairwise disjoint coalitions while in our model, the neighbourhoods of different agents may overlap.

 A few recent papers  \cite{ijcai2022p12,friedrich2023single,kanellopoulos2023tolerance,NS23},
like our work, are motivated by the observation that real-world agents 
can favour diverse neighbourhoods. Swap games with a single-peaked utility function were studied in \cite{ijcai2022p12}. The utility of an agent  increases monotonically with the fraction of
 same-type neighbours in the interval $[0, \lambda]$ for some $0<\lambda<1$, and decreases monotonically afterwards. Results on the existence of equilibria for specific classes of graphs for different ranges of $\lambda$, as well as tight bounds on the price of anarchy and stability with respect to the degree of integration defined in \cite{agarwal2020swap} were shown. 
Using the  single peak utility function for jump games, the authors of \cite{friedrich2023single} investigated the existence of equilibria. 
They showed that improving response cycles exist independently of the peak value,
 even for graphs with very simple structures. They also showed that while the existence of equilibria are not guaranteed  even on rings for $\lambda\geq 1/2$, there are  conditions under which they are guaranteed to exist, depending on the size  of the independent set, the number of empty nodes, and numbers of agents of each type set.
In addition, bounds on the price of stability and anarchy with respect to the degree of integration and some hardness results were given. 
Kanellopoulos et al. \cite{kanellopoulos2023tolerance} considered jump games with $k \geq 2$ types of agents and  an implicit ordering of the types.  The utility of an agent is affected by the distance between the types in the given ordering  whenever there are agents of those types in its neighborhood. 
In \cite{Narayanan2023diversityb}, the authors studied jump games in which the utility of an agent is the fraction of neighbours of a different type from itself.  It is shown that  to determine the existence of an equilibrium in the presence of stubborn agents is NP-hard, and the game is a potential game in regular graphs, and spider graphs, while there is always an equilibrium in trees.


\subsection{Organization of paper} \label{sec:org}
In Section~\ref{sec:Preliminaries}, we define our notation, utility functions and diversity measures more precisely, and make some preliminary observations.
Section \ref{sec:potential} shows our swap games are all potential games. Section
\ref{sec:Efficiency} and Section \ref{sec:special} present  results on the price of anarchy for the diversity measures for general graphs, and for special low-degree graphs respectively. 
Section~\ref{sec:experiments} presents our experimental results, and Section~\ref{sec:conclusions} presents some conclusions and discussion.


\section{Preliminaries} \label{sec:Preliminaries}

In this paper $G=(V, E)$ denotes a  connected graph where $V$, $E$ is the set of vertices, edges, respectively. Let $\Delta$ and $\delta$ denote the maximum and minimum degree of $G$, respectively. 
Let $N(v)$ denote the set of neighbours of vertex $v \in V$.
Let $\cT=\{1, 2, \ldots, t \}$ denote a set of {\em types}.
Let $\cX$ denote a set of {\em agents} that is partitioned into $t$ different types. For agent $A \in \cX$, we use $\tau(A)$ to denote its type. 
Throughout the paper we assume $|V| = |\cX| = n$. Let 
$k = \floor{n/t}$. The partition of agents into types is called {\em equitable} if every type has either $k$ or $k+1$ agents. Clearly, an equitable partition always exists for every $n$ and $t$. Sometimes we use the term \emph{equitable agents} to refer to a set of agents whose partition into types is equitable. 

An {\em assignment} of agents to vertices in $G$ is a bijection $L: \cX \rightarrow V$. 
We call $v = L(A)$ the {\em location} of agent $A$. Under an assignment $L$, we call agents $A$ and $B$ neighbors if $(L(A),L(B)) \in E(G)$, we call the edge $(L(A),L(B))$ \emph{monochromatic} if $\tau(A) = \tau(B)$, and {\em colorful} otherwise.  
For every agent $A$, an assignment $L$ naturally induces a type-vector $\Pi_{L,A} \in \N^t$ where $\Pi_{L,A}(i)$ denotes the number of agents of type $i$ located in $N(L(A))$. When $L$ is clear from the context, we abbreviate $\Pi_{L,A}$ as $\Pi_A$.
By an abuse of notation, we use $N(A) \subseteq \cX$ to denote  the set of neighboring agents of agent $A$, and $T(A) \subseteq \cT$ to denote the set of types of agents in $N(A)$. An agent $A$ is called {\em segregated} if $T(A) = \{\tau(A)\}$. 

Given an assignment $L$ for a graph $G$, a {\em utility function} $U$ is a real-valued function that maps an agent to its {\em utility} based on the agent's own type and the types of its neighboring agents in the graph. In Section~\ref{sec:utilities}, we define several specific utility functions.
The {\em swap game under a  utility function $U$} is defined as follows: Given an assignment $L$, a {\em move} in the game is a swap between two agents $A$ and $B$ such that both $A$ and $B$ increase their utilities by interchanging  their locations, leaving the locations of all other agents unchanged. Thus a move
yields  a new assignment.
An assignment is in an {\em equilibrium} with respect to $U$ if there is no move 
under $U$.  Given a utility function $U$, a graph $G$, and a set of agents $X$, we use $\eq(G, U)$ to denote the set of 
all equilibrium assignments for the swap game on $G$ under the utility function $U$.

A game on a graph $G$ is called   a {\em potential game} if and only if 
there is a non-negative real-valued function $\Phi$ 
on the set of assignments on $G$ such that $\Phi(L') <  \Phi(L)$ for any pair of assignments $L'$ and $L$ such that $L'$ is obtained from $L$ by a move in the game \cite{monderer1996potential}.

\subsection{Utility functions and diversity measures} \label{sec:utilities}

We define three utility functions to capture different notions of local diversity.

\begin{itemize}
\item $U_b$,  called {\em Binary}:  $U_b(x) =1$  if at least one neighbor of $x$ is of a different type than $x$, and
  is 0 otherwise.
    
\item $U_\#$, called {\em Difference-seeking} : $U_\#(x) $ is the number of  neighbors
  of $x$ whose type is different from $\tau(x)$.
    
\item $U_\tau$  called {\em Variety-seeking}:  $U_\tau(x) $ is the number of types that are different from $\tau(x)$ in the neighborhood of $x$.

\end{itemize}

$U_b$ is the simplest diversity-seeking utility function, while $U_\#$ and $U_\tau$ also consider the structure of types in the neighbourhood of an agent. Besides the simplicity and naturalness, these functions are also used in other studies on diversity. For example, $U_\#$ is a diversity-seeking variant of the similarity-seeking utility function in \cite{schelling-journal}, and a similar function was studied for jump games in \cite{NS23}.  $U_\tau$ corresponds to the biodiversity measure ``Numbers'' in \cite{bio_Nature}, i.e., ``the number of species in a site''. 
Note that segregated agents have utility $0$ in all three utility functions. Below are simple observations that follow directly from the definitions of the utility functions. 

\begin{observation} \label{obs:utility_observation}
    (1) If $t=2$, then $U_b = U_\tau$; \\
    (2) Every equilibrium assignment under any of $U \in \{U_b, U_\#, U_\tau\}$ has at most one type whose agents are segregated;\\
    (3) $\eq(G,U_\#), \eq(G,U_\tau) \subseteq \eq(G,U_b)$.
\end{observation}

Given an assignment $L$, every utility function $U$ induces a \emph{social welfare} $\sw(L,U) = \sum_{A \in \cX} U(A)$, a standard metric in algorithmic game theory. We consider the following four \emph{global} diversity measures, the first three of which correspond to the social welfare with respect to our utility functions. 

\begin{itemize}
    \item \emph{Degree of integration}: $\doi(L)$ is the fraction of agents $A$ with at least one neighbor of a type  different from $A$. We also consider two refined versions of $\doi$: 
    \begin{itemize}
        \item $\doic(L,j) = $ the fraction of agents $A$ having at least $j$ colorful edges incident to $A$; 
        \item $\doit(L,j) = $ the fraction of agents $A$ having at least $j$ distinct types different from $\tau(A)$ in its neighborhood. 
    \end{itemize}

    \item \emph{Colorful edges}: $\ce(L)$ is the number of colorful edges in $L$.

    \item \emph{Neighborhood variety:} $\nv(L)$ is the average number of types in the neighborhood of agents, where we only count types in the neighborhood that are different from the agent's type.

   \item \emph{Evenness}: $\ev(L) = \frac{1}{\sum_{A \in \cX} \norm{\Pi_A}_2^2}$.     
\end{itemize}

Note that 
    $\doi(L) = \sw(L,U_b)/n$, 
    $\ce(L) = \sw(L, U_\#)/2$, 
and
    $\nv(L) = \sw(L, U_\tau)/n$.
The term $\norm{\Pi_A}_2^2$ is used to measure local evenness of neighbours distribution of agent $A$.

We now briefly discuss why we choose these diversity measures. The degree of integration has been introduced in \cite{agarwal2020swap} to study the diversity in swap Schelling games.
The number of colorful edges is the number of heterophilic connections in a network, and is a simple measure of diversity. It is frequently used as a potential function, and was mentioned as a measure of segregation in the study of segregation \cite{antonio2004effects,BEL2016Gridanalysis}. 
Neighborhood variety is also a natural measure, and corresponds to the average number of species in a site in biodiversity. 
Evenness is another one of three biodiversity measures in \cite{bio_Nature}. Though evenness could be quantified in different ways such as by Shannon-Wiener index \cite{bio_Shannon}, we choose the $L^2$ norm because it is easier to estimate and has a similar property: $L^2$ is monotonically decreasing as the neighborhood becomes more evenly distributed. The $L^2$ measure has also been used in \cite{L2} to study diversity maximization with respect to graph coloring.

\subsubsection{Upper bounds on diversity measures}   \label{sec:opt_diversity}

The diversity measures above have the following trivial bounds: \\
(1) $\doi(L) \le 1$,
$\ce(L) \le |E(G)|$, and 
$\nv(L) \le  \min\{t-1, \Delta\}$;\\
(2) for $\delta$-regular graphs, 
$\ev(L) \le  t/ (n \cdot \delta^2)$, 
where we applied the Cauchy-Schwarz inequality 
    $\norm{\Pi_A}_2^2 = \sum_{i=1}^t \Pi_A(i)^2 \ge (\sum_{i=1}^t \Pi_A(i))^2 / t = \delta^2/t$.

\begin{figure}[ht!]
    \centering
   \includegraphics[scale=.8]{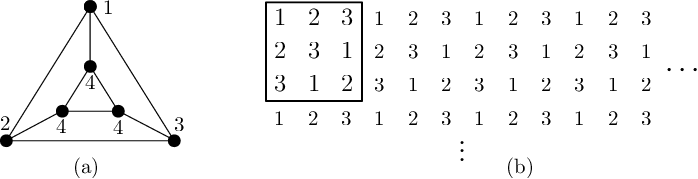}
   \caption{Non-achievable and achievable instances.}
    \label{fig:opt_diversity_example}
\end{figure}

We consider now  whether these bounds are always achievable by some assignment, for some set of agents (not necessarily equitable). 
Clearly, $\doi(L) = 1$ can be achieved on every graph for every $t\ge 2$, and the optimal assignment can be found with a greedy algorithm.
The condition $\ce(L) = |E(G)|$ is equivalent to  $L$ being a valid graph coloring of $G$, hence the equality is achievable if and only if $t \ge \chi(G)$. 
Furthermore, if the agents are equitable,  then $\ce(L) = |E(G)|$ corresponds to an equitable coloring of $G$. By Hajnal-Szemer{\'e}di theorem \cite{Haj_Sze_equitable_coloring}, every graph $G$ is equitable colorable with $t$ colors if $t \ge \Delta + 1$,  but this is not necessarily achievable if $t < \Delta + 1$. For $\nv(L)$, even assuming $G$ is $\delta$-regular for $\delta \ge t-1$, the upper bound $(t-1)$ is still not always achievable: see Figure \ref{fig:opt_diversity_example}-(a) for $t=4$, this example also implies that the upper bound of $\ev(L)$ cannot be achieved either. Nonetheless, on some graphs both optima can certainly be achieved, see  Figure \ref{fig:opt_diversity_example}-(b). We point out that the optimal assignment in  Figure \ref{fig:opt_diversity_example}-(b) is a tiling of the grid as marked in the illustration.

Finally,  $\doic(L,j)$ and $\doit(L,j)$ are both non-increasing for $j$. For $\delta$-regular graphs, $\doic(L,\delta)=1$ is equivalent to $\ce = |E(G)|$, $\doit(L,t-1)=1$ is equivalent to $\sw(L,U_\tau) = n(t-1)$. Hence, the previous discussion applies.

\subsection{Price of anarchy and stability}   \label{sec:PoA}

Consider a swap game on graph $G$ under utility function $U$. Let $\mu$ denote a diversity measure, $L^*$ be an assignment that {\em maximizes} $\mu$, and  $L^e$ and $L^m$ be \emph{equilibrium} assignments that {\em minimize} and {\em maximize} $\mu$ respectively. The price of anarchy (PoA) is defined to be 
    $\poa(\mu, U, G) = \mu(L^*) / \mu(L^e)$, and the price of stability is defined to be 
$\pos(\mu, U, G) = \mu(L^*) / \mu(L^m)$

For degree of integration, since at equilibrium it is possible that $\doic(L,j) = \doit(L,j) =0$, the PoA is not a suitable notion. Let $\wdoic(U,G,j)$ denote the minimal of $\doic(L,j)$ among all possible equilibrium $L$ under $U$ on graph $G$. Define $\wdoit(U,G,j)$ similarly. Clearly, $\wdoit(U,G,j) \le \wdoic(U,G,j)$, and both are non-increasing with $j$.

Let $f(k)$ and $g(k)$ be positive functions of $k$. We use $f(k)/g(k) \gtrsim p$  to mean that $\lim_{k \to \infty} f(k)/g(k) \ge p$. The notation $f(k)/g(k) \lesssim p$ and $f(k)/g(k) \eqas p$ are defined similarly. 

Observation \ref{obs:utility_observation} and the note after defining diversity measures imply the following.

\begin{corollary}   \label{cor:PoA}

(1) If $t=2$, then $\poa(\mu, U_b, G) = \poa(\mu, U_\tau, G)$;

    (2) $\poa(\mu, U_\#,G), \poa(\mu, U_\tau,G) \le \poa(\mu, U_b, G)$; 
    
    (3) $\poa(\sw, U_b,G) = \poa(\doi, U_b,G)$,  
    $\poa(\sw, U_\#,G) = \poa(\ce, U_\#,G)$, and 
    $\poa(\sw, U_\tau,G) = \poa(\nv, U_\tau,G)$;
    
    (4) $\poa(\ce, U,G) \le \delta/(\wdoic(U,G,1) + \wdoic(U,G,2))$.
\end{corollary}

\begin{proof}
    (1) and (2) follow directly from Observation \ref{obs:utility_observation}. (3) follows by our earlier observation that for $U_\#$, $\sw(L,U_\#) = 2\ce(L)$ on every $G$ and $L$. For (4), fix $U$, let $w_j = \wdoic(U,G,j)$. Then, for equilibrium assignment $L$ we have
        $\ce(L) \ge n (w_1 - w_2) \cdot 1/2 + n w_2 \cdot 2/2$. 
    Hence, 
        $\poa(\ce, U,G) \le (n\cdot \delta/2) / \ce(L) 
        \le \delta/(w_1 + w_2)$.
\end{proof}

For the refined degree of integration measures, rather than defining the price of anarchy, we are interested in the worst-case over all equilibria. For any $j$, and a given utility function $U$, define 
$$\wdoic{j} = \min_{L \in \eq(G, X,  U)} \{ \doic{L}{j} \} \mbox{~~~ and ~~~} \wdoit{j} = \min_{L \in \eq(G, X, U)}  \{ \doit{L}{j} \}$$


\section{Existence of equilibrium}
\label{sec:potential}

In this section, we show that swap games under our utility functions are potential games. Before  this, we determine the conditions under which swaps can occur under our utility functions. Throughout  this section we assume agents are equitable.

Clearly, under the  utility function $U_b$,  a swap 
can occur between two agents if and only if  they are  segregated and of different types. Next we turn our attention to $U_\#$.

\begin{lemma}   \label{lem:U_sharp_basic_property} 
    Let $A$ and $B$  denote agents of different types in an equilibrium assignment under $U_\#$. Then 
    (1) $U_\#(A) + U_\#(B) \ge \delta$;
    (2) If 
        $U_\#(A) = U_\#(B) = c$,
     then, either $N(A)$ contains at least $\delta - c$ agents of type $\tau(B)$, or $N(B)$ contains at least $\delta - c$ agents of type $\tau(A)$;
     (3) If $A$ and $B$ are neighbors, $U_\#(A) + U_\#(B) \ge \delta + 1$.
\end{lemma}

\begin{proof}
    (1) Let $\tau(A) = i, \tau(B) = j$ and assume $U_\#(A) = r$. Then, $N(A)$ contains $\ge \delta-r$ agents of type $i$. If $U_\#(B) \le \delta - r - 1$, then $N(B)$ contains $\ge r+1$ agents of type $j$. But then $A$ and $B$ could swap, a contradiction to the equilibrium. 

    (2) Assume otherwise, then, $N(A)$ contains
        $\ge \delta - (\delta - c - 1) = c+1$
    agents of type different from $\tau(B)$, and $N(B)$ contains 
        $\ge c+1$
    agents of type different from $\tau(A)$. Hence, $A$ and $B$ could swap, a contradiction.

    (3) Let $U_\#(A) = i$ and $U_\#(B) = j$, and for the sake of a contradiction assume $i+j \le \delta$. Then, $N(A)$ contains exactly $\delta - i \ge j$ many agents of type $\alpha$, and similarly, $N(B)$ contains exactly $\delta - j \ge i$ many agents of type $\beta$.  Then $A$ and $B$ would increase their utilities by swapping - $A$ at its new location would have at least $i+1$ many agents of type $\beta$, that is, the old neighbors of $B$ of type $b$ as well as $B$ itself, and therefore utility at least $i+1$, and similarly $B$ at its new location would have utility at least $j+1$ - a contradiction to the equilibrium. 
\end{proof}

\begin{corollary}   \label{cor:U_sharp_num_of_types}
    In an equilibrium under $U_\#$, the number of types that can have agents with exactly $c$ colorful edges is
        $\le 2 c/(\delta - c)  +1$.
\end{corollary}

\begin{proof}
    Suppose there are $q$ types, wlog assume they are $\{1,\ldots,q\}$, and let $A_1, \ldots, A_q$ be $q$ agents such that $\tau(A_i) = i$. The condition says $U_\#(A_i) = c$ for every $i$. For each $A_i$, we call an agent in $N(A_i)$ whose type is different from $i$ a \emph{spot}. Hence, there are exactly $qc$ spots in total. (2) in Lemma \ref{lem:U_sharp_basic_property} says that for every pair $A_i$ and $A_j$, at least $\delta-c$  spots must be taken. Note that for distinct pairs, these clusters of $\delta-c$ spots are non-overlapping. This implies
        ${q \choose 2} (\delta - c) \le q c$,
    giving the desired bound.
\end{proof}

We proceed to study $U_\tau$. 
\begin{lemma} \label{lem:swap-condition-U_tau} 
    Let $A$ and $B$ be two agents of different types, 
    let $T(A), T(B)$ denote the set of types in the neighborhood of agent $A$ and $B$, respectively, before a possible swap. Under $U_\tau$, $A$ and $B$ have the incentive to swap if and only if: 
    (i) $A$ and $B$ are not neighbors, 
    (ii) $U_\tau(A) = U_\tau(B)$,  and 
    (iii) $\tau(A) \in T(A) -  T(B)$ and $\tau(B) \in  T(B) - T(A)$. 
\end{lemma}

\begin{proof}
    The ``if'' part is easy to check. We thus show the ``only if'' part. Assume $A$ and $B$ have the incentive to swap. 
    
    (i) Assume $A$ and $B$ are neighbors, i.e., 
        $A \in N(B)$ and $B \in N(A)$.
    Before the swap, 
        $U_\tau(A) = |T(A) - \tau(A)|$,
    and after the swap,
        $U_\tau(A) = |T(B) - \tau(A) - \tau(B)| + 1$.
    Hence, for $A$ to swap one needs
        $|T(B) - \tau(A) - \tau(B)| + 1 > |T(A) - \tau(A)|$.
    Because $A \in N(B)$, we have
        $|T(B) - \tau(A) - \tau(B)| = |T(B) - \tau(B)| - 1$.
    Combining these two we get
        $|T(B) - \tau(B)| > |T(A) - \tau(A)|$.
    Symmetrically, for $B$ to have the incentive to swap one needs
        $|T(A) - \tau(A)| > |T(B) - \tau(B)|$, 
    a contradiction. 

    (ii) By (i), $A$ and $B$ are not neighbors, hence, the swap would make $A$ and $B$ inherit each other's open neighborhood. Hence, $A$ at its new location would have utility at most
        $U_\tau(B)+1$. 
    For $A$ to swap,  we must have
        $U_\tau(B)+1 > U_\tau(A)$,
    i.e.,
        $U_\tau(B) \ge U_\tau(A)$.
    By symmetry,
        $U_\tau(A) \ge U_\tau(B)$,
    this proves (ii).

    (iii) We already know $A$ and $B$ are not neighbors and (ii) holds, the only way that $A$ would have incentive to swap with $B$ is that it has utility exactly 
        $U_\tau(B) + 1$
    at its new location, this implies that
        $\tau(A) \not\in T(B)$ and $\tau(B) \in T(B)$.
    Similarly, for $B$ to increase its utility by swapping, 
        $\tau(B) \not\in T(A)$ and $\tau(A) \in T(A)$.
    This proves (iii).
\end{proof}

\begin{corollary} \label{cor:U_tau_property}
    Let $A$ and $B$ be two agents at equilibrium under $U_\tau$.
    Suppose $U_\tau(A) = U_\tau(B)$, and $\tau(A) \in T(A)$, $\tau(B) \in T(B)$. Then,  either $\tau(A) \in T(B)$ or $\tau(B) \in T(A)$.
\end{corollary}

Let $T_s = \{i \in \cT: \tau(A)=i, U_\tau(A) = s, \tau(A) \in T(A)\}$, i.e., the set of types of nodes who have $U_\tau$ utility value exactly $s$ which moreover have a neighbor of the same type.

\begin{lemma}   \label{lem:U_tau_bounds_on_Ts}
    $|T_s| \leq 2s+1$.
\end{lemma}

\begin{proof}
    Let $|T_s| = q$, and $A_i, A_j$ be two agents of type $i,j$, respectively, and suppose they both satisfy the condition defining $T_s$, i.e.,
        $U_\tau(A_i) = U_\tau(A_j) = s$,
    and $i \in T(A_i)$, 
        $j \in T(A_j)$.
    Then, $A_i, A_j$ satisfies the condition for Corollary \ref{cor:U_tau_property}, hence,
        $i \in T(A_j)$
    and $j \in T(A_i)$.
    Since these pairwise constraints must be satisfied for every pair from the $q$ types, the same argument in Corollary \ref{cor:U_sharp_num_of_types}  implies ${q \choose 2} \leq qs$, giving the desired bound.
\end{proof}

\begin{corollary} \label{cor:U_tau_num_of_types}
     In an equilibrium under $U_\tau$, for $1 \leq c \le \delta-1$, the number of types that can have agents with exactly $c$ colorful edges is $\le c^2 + 2c$.
\end{corollary}

\begin{proof}
    Let $A$ be an agent of interest to the question. Then, $c\ge 1$ implies $1 \le U_\tau(A) \le c$, and $c \le \delta-1$ implies $\tau(A) \in T(A)$. 
    Let $T_{c, s} \subseteq T_s$ be the subset of types that can have agents with exactly $c$ colorful edges, besides the condition for $T_s$. Then, the number of types in question is 
        $|\cup_{s=1}^c T_{c, s}| 
        \le \sum_{s=1}^c |T_{c, s}|
        \le \sum_{s=1}^c |T_{s}|$.
    By Lemma~\ref{lem:U_tau_bounds_on_Ts}, this is at most
        $\sum_{s=1}^c (2s + 1) = c^2+2c$.
\end{proof}

We are now ready to show that all three swap games we study are potential games. We define  $\Phi(L)$ to be the number of monochromatic edges in assignment $L$ and show that $\Phi$ is a potential function for all three utility functions that we consider. 
Similar potential functions were used previously, eg. \cite{schelling-journal,chauhan2018schelling,echzell2019convergence}.

\begin{theorem} \label{thm:potential_game}
    On every graph $G$, the swap games under $U_b, U_\tau, U_\#$  are all potential games that reach their respective equilibria after at most $|E|/2$ moves.
\end{theorem}

\begin{proof}
       Fix and assignment $L$ and suppose $A$ and $B$ are two swapping agents of different types, at location $v$ and $w$ respectively in $L$. Let $L'$ be the assignment after $A$ and $B$ swap.  Observe that the swap of $A$ and $B$ affects only edges incident to $v$ and $w$. We show that $\Phi(L')\leq \phi(L)-2$, implying that equilibrium is reached after at most $|E|/2$ steps. 
    
    For $U_b$, since each swap reduces the number of segregated agents by 2, we have $\Phi(L') \leq \phi(L) - 2$. In fact, the game under $U_b$ reaches an equilibrium in at most $|V|/2$ steps.

    For $U_\#$, by definition, $A$ and $B$ swap if and only if they both increase the number of colorful edges after the swap incident on them; hence $\phi(L') \leq \phi(L)-2$. 

    For $U_\tau$, Lemma~\ref{lem:swap-condition-U_tau} implies that $\tau(A) \in T(A) -  T(B)$ and $\tau(B) \in  T(B) - T(A)$. Observe that $\tau(A) \in T(A)$ implies that before the swap, there is at least one monochromatic edge incident to $v$ in $L$, and $\tau(B) \not\in T(A)$ implies that after the swap there is no monochromatic edge incident to $v$ in $L'$. The same phenomenon holds for $w$. Hence, $\phi(L') \leq \phi(L)-2$. 
\end{proof}

\section{Efficiency at equilibrium}
\label{sec:Efficiency}

In this section, we study the price of anarchy and stability with respect to the four diversity measures we introduced in Section~\ref{sec:Preliminaries}.
\footnote{For the diversity measures and utility functions we study, the optimal assignment can be seen to be an equilibrium, therefore the price of stability is 1.}. 
We will be mainly interested in the behaviour of PoA  when $k\to \infty$, without loss of generality  we assume $k$ is always admissible with respect to appearing constraints (e.g., $k$ is a multiple of $(t-1)$, etc).  
We first discuss the case $t \ge 3$, since the case $t=2$  has a different behaviour. Hence,  $t\ge 3$ is assumed unless otherwise specified.

\begin{figure}[ht!]
    \centering
   \includegraphics[scale=.5]{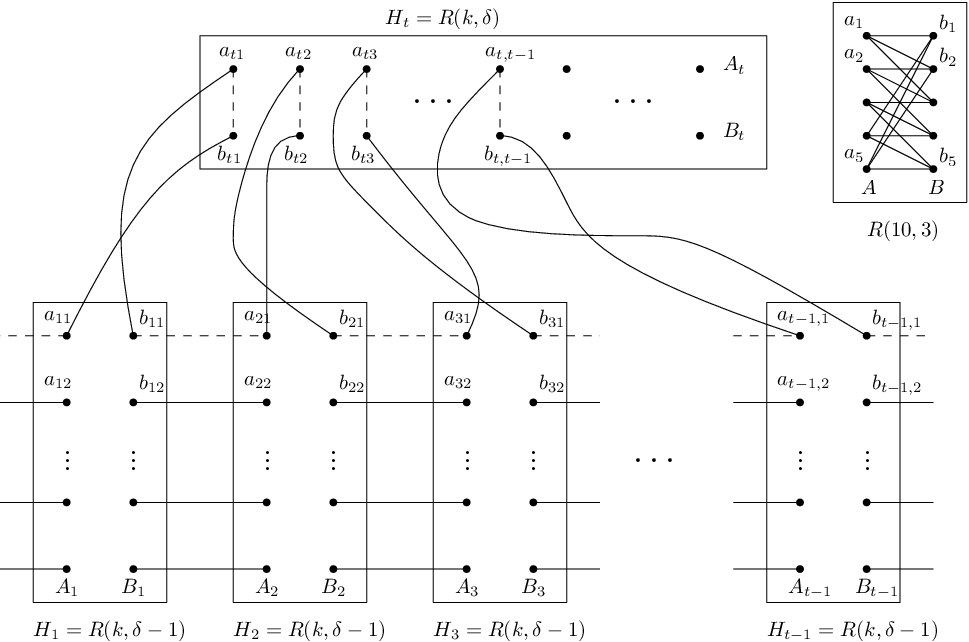}
    \caption{The $\delta$-regular graph $G^*$ for $t\ge 3$.} 
    \label{fig:G_star}
\end{figure}

We first construct a regular graph that will be used for proving lower bounds on PoA. Let $R(p,\delta)$ denote the bipartite $\delta$-regular graph on $p=2q$ vertices constructed as follows: name the vertices in the two parts respectively by $a_1, a_2, \ldots, a_q$ and $b_1, b_2, \ldots, b_q$, for every $i$ connect $a_i$ to $b_i, b_{i+1}, \ldots, b_{i+\delta-1}$, where the index is calculated mod $q$. See an illustration of $R(10,3)$ in the top right corner of Figure \ref{fig:G_star}. We use $R(k,\delta)$ as gadgets to construct a $\delta$-regular graph $G^*$  on $n=kt$ vertices,  as shown in Figure \ref{fig:G_star}, for $t\ge 3$, as follows. Take $t-1$ copies of $R(k, \delta-1)$, name them as $H_1, \ldots, H_{t-1}$, whose two parts of vertices are denoted by $A_i,B_i$, respectively. Connect $B_{i}$ to $A_{i+1}$ via a perfect matching except without the edge between $b_{i1}$ and $a_{i+1,1}$, shown as dashed line in Figure \ref{fig:G_star}. Note that $B_{t-1}$ is connected with $A_1$. Take $H_t = R(k,\delta)$, and remove its $t-1$ edges $(a_{ti},b_{ti})$ for $i=1,\ldots, t-1$. Finally, connect $b_{i1}$ to $a_{ti}$, and connect 
$a_{i1}$ to $b_{ti}$, for every $i=1,\ldots,t-1$. 
In Figure \ref{fig:G_star}, in order to avoid a crowd of edges, note that the edges between $A_i$ and $B_i$ are not shown (but they should be in $G^*$ as specified). It is easy to see that $G^*$ is $\delta$-regular.

The theorem below shows that the simple upper bounds on $\doi$, $\nv$ and $\ev$ given in Section~\ref{sec:opt_diversity} are all tight for equitable agents. Interestingly, these upper bounds can be \emph{simultaneously} achieved via a single equilibrium assignment on a single graph $G^*$. Roughly, this equilibrium has segregated agents of one type, and each of the agents of other types is surrounded by agents of another type. Hence, although this equilibrium has many colorful edges, it has very limited variety and evenness. 

\begin{theorem} \label{thm:PoA_ev_etal}
    (1) For equitable agents,  for every $U \in \{U_b, U_\#, U_\tau\}$, for every graph $G$,
        $\poa(\doi, U, G) \lesssim
        t/(t-1)$, 
        $\poa(\sw, U_b, G) \lesssim
        t/(t-1)$,
        $\poa(\nv, U, G) \le t$, 
        $\poa(\sw, U_\tau, G) \le t$, 
    and for every $\delta$-regular graph $G$,
        $\poa(\ev, U, G) \le t$.
        
    (2) There exists an assignment on $G^*$ that is simultaneously an equilibrium under  $U_b, U_\#$ and $U_\tau$, and achieves all upper bounds in (1), provided $t$ divides $\delta-1$, $k = \omega(\delta^2)$, and both $k$ and $\delta$ are sufficiently large.
    
(3) For equitable agents,  for every $U \in \{U_b, U_\#, U_\tau\}$, 
        $\pos(\doi, U, G^*) = \pos(\nv, U, G^*) = \pos(\ev, U, G^*) = 1$.    
\end{theorem}

\begin{proof}
    (1) For $\doi$, the maximum is $1$, and it follows from Observation \ref{obs:utility_observation}-(2) that in any equilibrium, $\doi \ge k(t-1)/(kt)$. 
    For $\nv$, the maximum is $t-1$, and it follows from Observation \ref{obs:utility_observation}-(2) that in any equilibrium, $\nv \ge k(t-1)/n$. 
    The bounds for $\poa(\sw, U_b, G)$ and $\poa(\sw, U_\tau, G)$ follows by combining these with Corollary \ref{cor:PoA}.
    For $\ev$, in Section \ref{sec:opt_diversity} we discussed that the maximum is
        $t / (n \cdot \delta^2)$.
    On the other hand, trivially 
        $\ev(L) \ge 1/ (n \cdot \delta^2)$
    for equilibrium $L$ (in fact for all $L$). This implies the stated upper bounds on PoA. 
    
    (2) To show the tightness of the above upper bounds, we use the regular graph $G^*$ in Figure \ref{fig:G_star}.  We first give an optimal assignment $L^*$. We  describe an assignment for $H_1$,  the same strategy is used for other $H_i$'s. Let $r=(\delta-1)/t$. Assign to the  vertices $a_{11}, a_{12}, \ldots$ in $A_1$ successively,  $r$ many agents of type $1$ followed by $r$ many agents of type  $2$, etc, and restart when $t$ types are used. Assign agents to the vertices in $B_1$ in a similar way, except for starting with type 2. It is easy to see that different types of agents have equal size.  Note that there is no segregated agent, hence, $\doi(L^*) = 1$. Also, every agent has all $t$ types in its neighborhood, hence, $\nv(L^*) = t-1$. 
    Observe that all vertices have an evenly distributed neighborhood consisting of  $r+1$ agents of type $i+1$ (respectively $i-1$) for vertices $a_{ij}$ (respectively $b_{ij}$) and $r$ agents of every other type.
    Note that the local evenness  for $a_{ij}$ is $(t-1)r^2 + (r+1)^2 = tr^2 + O(r) \le \delta^2/t + O(\delta/t)$, and the same upper bound holds for $b_{ij}$. Hence, $\ev(L^*) \ge 1/ (kt \cdot (\delta^2/t + O(\delta/t))) $.

    Next, we give an equilibrium assignment $L^e$. We first describe the assignment of $H_1$. Partition vertices in $A_1$ and $B_1$ equally into $t-1$ types, successively assign agents of type $1, 2, \ldots, t-1$ the $t-1$ parts in $A_1$, and agents of type $2, \ldots, t-1, 1$ to the corresponding parts in $B_1$. Apply the same assignment for every $H_i$ for $i\le t-1$.  Finally, assign agents of type $t$ to $H_t$. It is easy to see that different types of agents have equal size. Observe that except edges inside $H_t$, all edges are colorful, this implies $L^e$ is at equilibrium for $U_\#$, hence it is also at equilibrium for $U_b$ by Observation \ref{obs:utility_observation}-(3). Also, except $H_t$, no agent $A$ satisfies $\tau(A) \in T(A)$, hence, Lemma \ref{lem:swap-condition-U_tau} implies that $L^e$ is also at equilibrium for $U_\tau$. 
    Note that except $O(t)$ agents in $H_t$, the rest of the agents in $H_t$ are segregated. Hence, $\doi(L^e) \le (k(t-1) + O(t))/n$. 
    Note that, for $1 \le i\le t-1$ every $H_i$ has $O(t\delta)$ agents with $U_\tau = 2$, the rest all having $U_\tau = 1$, and agents of $H_t$ all have $U_\tau \le 1$. This implies $\nv(L^e) \le (k(t-1) + O(t^2 \delta)) / n$. 
    Similarly, since except $O(t^2 \delta)$ many agents, all other agents have only one type in their neighborhood, this implies  $\ev(L^e) \le  1/ ((kt-O(t^2\delta)) \cdot \delta^2)$. 
    
    The asymptotic lower bounds of PoA can be directly calculated from the discussion of the above two assignments, provided 
     $k = \omega(\delta^2)$ and both $k$ and $\delta$ are sufficiently large.

(3) Follows from the fact that the assignment $L^*$ in (1) above is an equilibrium for all three utility functions. 
\end{proof}

Under $U_\#$ or $U_\tau$, the equilibrium structures on colorful edges we explored in Section \ref{sec:potential} enable us to show nontrivial upper bounds. 

\begin{theorem} \label{thm:PoA_ce}
    For every graph $G$ and equitable agents,
    
    (1) $\poa(\ce, U_b, G) \lesssim \Delta t/(t-1)$, and this is tight on $G^*$; also $\pos(\ce, U_b, G^*) = 1$;
        
     (2) $\poa(\ce, U_\#, G) = \poa(\sw, U_\#, G) \lesssim 2 \Delta/\delta$, furthermore: 
     
     $\quad$ (i) $\poa(\ce, U_\#, G) \to \Delta/\delta$, for $t \ge \omega(\delta \log \delta)$, 
     
     $\quad$ (ii)  $\poa(\ce, U_\#, G) \lesssim 2/(\delta/\Delta + \Omega(t^2/\delta \Delta - 1/\Delta))$, for $\Omega(\sqrt{\delta})\le t \le O(\delta)$;

    (3) $PoA(\ce, U_\tau, G) \lesssim \Delta t/(t-1)$, furthermore: 
    
    $\quad$ (i) $\poa(\ce, U_\tau, G) \to \Delta/\delta$, for $t \ge \omega(\delta^3)$, 
    
    $\quad$ (ii) $PoA(\ce, U_\tau, G) \le O(\Delta/t^{1/3})$, for $t \le O(\delta^3)$;
    
    (4) $\pos(\ce, U_\#, G) = 1$ when $t \geq \Delta+1$.
\end{theorem}

\begin{proof}
    (1) We have $\ce \le kt \cdot \Delta/2$, and Observation \ref{obs:utility_observation}-(2) implies $\ce \ge k(t-1)/2$ (note that every colorful edge can be shared by two agents), which implies the upper bound.
 
    For the lower bound on PoA, again we use the regular graph $G^*$ in Figure \ref{fig:G_star}. We first give an optimal assignment $L^*_b$: assign agents of type 1 to $A_t$, type 2 to $B_t$. and then successively assign to  $A_1, B_1, A_2, B_2, \ldots$  agents of type  $3,4,5,6,\ldots$, until $t$ types have been used, then restart from type $1$, until it finishes with type $t$ on $B_{t-1}$. 
    Observe that only edges from $H_i$ (for $1\le i\le t-1$) to $H_t$ are possible monochromatic edges, which is at most $O(t)$. Hence, $\ce(L^*_b) \ge kt\cdot \delta/2 - O(t)$. 

    Next, we give an equilibrium assignment $L^e_b$: assign agents of type $i$ to $H_i$  for $i=1,\ldots,t$. It is easy to check that this is an equilibrium under $U_b$. Note that colorful edges are shared between $H_i$ and $H_{i+1}$ for $i=1,\ldots, t-1$ mod $t-1$, and between $H_i$ and $H_t$. Hence, $\ce(L^e_b) \le k(t-1)/2 + O(t)$. This together with $\ce(L^*_b)$ implies the lower bound on PoA. Finally, the fact that $L^*_b$ is an equilibrium implies that $\pos(\ce, U_b, G^*)=1$.  
    
    (2) Since $\poa(\ce, U_\#, G) = \poa(\sw, U_\#, G)$, we work with $\sw$ instead.  Trivially, $\sw(L,U_\#) \le n\cdot \Delta$, hence it suffices to show at equilibrium $\sw(L,U_\#) \gtrsim n\cdot \delta/2$. This is true if every agent has utility at least $\delta/2$. If not, suppose some agent has utility $r < \delta/2$, by (1) of Lemma \ref{lem:U_sharp_basic_property}, all agents of other types have utility at least $\delta - r$. Hence, 
        $\sw(L,U_\#) \ge kr + (t-1)k (\delta - r) \ge tk \cdot \delta / 2 \gtrsim n\cdot \delta/2$.

    For the ``furthermore'', let $\delta/2 \le x \le \delta-1$, let $L$ be an equilibrium assignment under $U_\#$. Let $f(x)$ denote the number of types that can have agents with $\le x$ colorful edges. Then, earlier we have observed that at most one type can have agents of utility $<\delta/2$, hence, by Corollary \ref{cor:U_sharp_num_of_types}, 
    $f(x) \le 1+ \sum_{c=\delta/2}^x (2c/(\delta-c) + 1)$, estimating by integration, this is
    \begin{align*}
        f(x) &\le 2\int_{\delta/2}^{x+1} c/(\delta-c) dc  +(x+1-\delta/2)\\
        &= 2\delta \ln \frac{\delta}{2(\delta-x-1)} - (x + 1 -\delta/2) = g(x,\delta), 
    \end{align*}
    this upper bound holds for $x \le \delta - 2$. 

    For (i): again by Corollary \ref{cor:U_sharp_num_of_types}, the number of types that can have agents of utility $\delta-1$ is at most
        $2(\delta-1)+1$.
    Hence, 
        $f(\delta-1) \le g(\delta-2,\delta) + 2(\delta-1)+1 \le O(\delta \log \delta)$.
    Hence, $\sw(L,U_\#)  \ge k(t- f(\delta-1) ) \cdot \delta \ge k(t-O(\delta \log \delta))) \cdot \delta \to kt\cdot \delta$, for $t = \omega(\delta \log \delta)$.

    For (ii): when $\delta > t$, let $x+1 = y\delta$, then 
        $g(x,\delta)= 2\delta \ln (1/(2-2y))  - (y-1/2)\delta$.
    One can directly check that when 
        $y=1/2 + ((t-1)/\delta)^2$ 
    or equivalently when
        $x = \delta/2 + (t-1)^2 / \delta - 1$, 
    we have
        $f(x) \le g(x,\delta) \le t-1$ holds as long as $\delta \ge 4 t$.
    In this case, we have
        $\sw(L,U_\#) \ge k(t-g(x,\delta)) \cdot x + k (g(x,\delta) - 1) \cdot \delta/2
        \ge kt \cdot \delta/2 + kt\cdot \Omega(t^2/\delta - 1)$.
    This implies the bound of PoA.

    (3) The upper bound follows by (2) of Corollary \ref{cor:PoA} and the upper bound of $U_b$ from (1). For the ``furthermore'', let $f(x)$ denote the same as in the proof of (2). By Corollary \ref{cor:U_tau_num_of_types},
        $f(x) \le 1 + \sum_{1\le c \le x} (c^2 + 2c) \le x^3$
    for $x\ge 3$. Choose $x$ to be an integer, then agents of all other types have $\ge x+1$ colorful edges. Hence, 
        $\ce(L) \ge k (t-x^3) \cdot (x+1)/2$.
    Then, (i) and (ii) follow by letting $x=\delta-1$ and $x=\Theta(t^{1/3})$, respectively.

    (4) Since the agents are equitable,   $\ce(L) = |E(G)|$ corresponds to an equitable coloring of $G$. By Hajnal-Szemer{\'e}di theorem \cite{Haj_Sze_equitable_coloring}, every graph $G$ is equitable colorable with $t$ colors if $t \ge \Delta + 1$. Since all agents have maximum utility, this coloring corresponds to an equilibrium assignment, which implies that $\pos(\ce, U_\#, G)= 1$.
    \end{proof}

Next we prove bounds on the worst case degree of integration. 

\begin{theorem} \label{thm:PoA_DOI}
    For every graph $G$ and equitable agents, for every $U\in \{U_b,U_\#, U_\tau\}$, $\wdoic(U,G,1) \ge \wdoit(U,G,1) \gtrsim 1- 1/t$. Furthermore,
    
    (1) $\wdoit(U_b,G^*,2) \le \wdoic(U_b,G^*,2) \lesssim 0$; 

    (2) $\wdoic(U_\#,G,\delta/2) \gtrsim 1 - 1/t$, 
        and $\wdoic(U_\#,G,j) \gtrsim 1 - 2\delta (j-\delta/2)/t$ for $\delta/2 + 1 \le j \le \delta-1$;

    (3) $\wdoic(U_\tau,G,2) \gtrsim 1 - 4/t$, $\wdoic(U_\tau,G,3) \gtrsim 1 - 12/t$, and $\wdoic(U_\tau,G,j) \gtrsim 1 - (j-1)^3/t$ for $j\ge 4$;

    (4) there exists an assignment $L$ on $G^*$, that is an equilibrium under both $U_\#$ and $U_\tau$, such that
            $\doit(L,2) = 0$ and $\doic(L,\delta) = 1$.
        In particular, 
            $\wdoit(U_\#,G^*,2) = \wdoit(U_\tau,G^*,2) \eqas 0$.
\end{theorem}

\begin{proof}
    The lower bound follows by Observation \ref{obs:utility_observation}-(2).

    (1) For $U_b$,  combine (4) of Corollary \ref{cor:PoA} and (1) of Theorem \ref{thm:PoA_ce}, on $G^*$ we have
        $\delta t/(t-1) \lesssim \delta / (\wdoic(U_b,G^*,1) + \wdoic(U_b,G^*,2))$.
    Hence,
        $\wdoic(U_b,G^*,1) + \wdoic(U_b,G^*,2) \lesssim (t-1)/t = 1 - 1/t$.
    Since
        $\wdoic(U_b,G^*,1) \gtrsim 1- 1/t$,
    it follows that 
        $\wdoic(U_b,G^*,2) \lesssim 0$. 

    (2) By (1) of Lemma \ref{lem:U_sharp_basic_property}, at equilibrium under $U_\#$ at most one type can have agents with $<\delta/2$ colorful edges in their neighborhood, this implies the lower bound of $\wdoic(U_\#,G,\delta/2)$. For larger $j$, recall the estimation of $f(x)\le g(x,\delta)$ in (2) of Theorem \ref{thm:PoA_ce}, let $x+1 = \delta/2 + z$, apply $\ln x \le x-1$, a simple calculation gives
        $g(x,\delta) \le z(3\delta + 2z)/(\delta-2z) \le 2 \delta z$
    when $\delta/2 \le x \le \delta-2$. Let $x=j-1$ be an integer, this implies the number of types whose agents all have $\ge x+1=j$ colorful edges is at least
        $t- f(x) \ge t-2\delta z = t- 2\delta (j-\delta/2)$.
        
    (3) We have 
        $\wdoic(U_\tau,G,j+1) \ge (t-f(j))/t$,
    the bound can be calculated via the estimation of $f(x)$ in (3) of Theorem \ref{thm:PoA_ce}.

    (4) The desired $L$ is the assignment given in (2) of Theorem \ref{thm:PoA_ev_etal}.   
\end{proof}

\subsection{The case $t=2$}  \label{sec:PoA_t_is_2}

Recall, by Observation \ref{obs:utility_observation}, that $U_\tau = U_b$ when $t=2$. We can obtain better bounds of PoA for $U_b$, e.g.,
    $\poa(\sw, U_b, G) \lesssim 2/(1+1/\Delta)$,
    $\poa(\ce, U_b, G) \lesssim \Delta$, etc.
Indeed, suppose wlog that type-1 has segregated agents, then, for every type-2 agent to have at least one colorful edge, at least $k/\Delta$  type-1 agents also have at least one colorful edge.  This implies the claimed bound for $\sw$. For $\ce$, note that in this case we have $\ce \ge k$ (instead of $k/2$), which implies the claimed bound. One can also show that these bounds are tight, we omit the details.

For $U_\#$, the results for $t\ge 3$ still apply to $t=2$. Furthermore, for equitable agents and $\delta$-regular graphs, unlike the other two utility functions, it is easy to see that there is at most one segregated agent. Indeed, by Lemma \ref{lem:U_sharp_basic_property} -(1), the existence of one segregated agent implies agents of the other type have the maximum number of colorful edges, and a simple counting argument proves the claim.  It follows that  
$PoA(\doi, U_\#, G)\eqas 1$ and $PoA(\nv, U_\#, G) \eqas 1$. It can be shown that 
$PoA(\ce, U_\#, G) \approx 2$; for the upper bound, the argument for $t \geq 3$ still applies, and for the lower bound follows from a specific regular graph.

\section{Cycles, cylinders, tori}
\label{sec:special}

Let $C_n$ denote the cycle on $n$ vertices, 
and $P_n$ denote the cylinder on $n$ vertices, i.e., a $3$-regular graph that is of a rectangle form which has $2$ rows and $n/2$ columns.
Let $T_n$ denote the $4$-regular torus on $n$ vertices that corresponds to a $\sqrt{n} \times \sqrt{n}$ grid.
Cycles and grids have been extensively studied for Schelling games as they model residential neighborhoods. In this section we study the PoA with respect to $\sw$ and $\ce$ and obtain tight or better bounds than those implied from Section \ref{sec:Efficiency}. We also study cylinders because they provide a  smooth transition from cycles to tori, besides, cylinders also model some residential neighborhoods. Table \ref{tab:cylcles_cylinders_grids} summarizes our results.

\begin{table}[ht!]
\caption{{PoA for cycles, cylinders, and tori. Except the PoA of $\ce$ for $U_\#$ on cylinders and tori, all other bounds are tight. The PoA of $\ce$ for $U_\#$ on $T_n$ lies in-between 
$t/(t-1)$ and $\min\{ t/(t-13/4), 4t/(3(t-1))\}$. A bold font represents a better bound improving the corresponding bound in Section \ref{sec:Efficiency}, while the non-bold matches the corresponding bounds.
} }
\label{tab:cylcles_cylinders_grids}
\centering
\begin{tabular}{|l|l|l|l|l|l|}
\hline
$t$                      &                        &          & Cycles      & Cylinders             & Tori         \\ \hline
\multirow{4}{*}{$2$}     & \multirow{2}{*}{$\sw$ }                 & $U_b, U_\tau$    & $4/3$       & $3/2$                 & $8/5$                \\ \cline{3-6}
                         &                        & $U_\#$   & {\bf 3/2}       & {\bf 3/2}                 & {\bf 5/3}                \\ \cline{2-6}
                         & \multirow{2}{*}{$\ce$} & $U_b,U_\tau$    & $2$         & $3$                   & $4$                  \\ \cline{3-6} 
                         &                        & $U_\#$   & {\bf 3/2}       & {\bf 3/2}                 & {\bf 5/3}                \\ \hline
\multirow{6}{*}{$\ge 3$} & \multirow{3}{*}{$\sw$} & $U_b$    & $\frac{t}{t-1}$   & $\frac{t}{t-1}$             & $\frac{t}{t-1}$            \\ \cline{3-6}
                        &                        & $U_\#$   & {$\mathbf{\frac{t}{t-1}}$}     & {$\big[\mathbf{\frac{t}{t-1},\frac{t}{t-7/3}} \big]$}  & {\bf see caption}                \\ \cline{3-6} 
                         &                        & $U_\tau$ & $\frac{2t}{t-1}$  & $\min\{t,\frac{3t}{t-1}\}$  & $\min\{t,\frac{4t}{t-1}\}$ \\ \cline{2-6} 
                         & \multirow{3}{*}{$\ce$} & $U_b$    & $\frac{2t}{t-1}$  & $\frac{3t}{t-1}$            & $\frac{4t}{t-1}$           \\ \cline{3-6} 
                         &                        & $U_\#$   & {$\mathbf{\frac{t}{t-1}}$}     & {$\big[\mathbf{\frac{t}{t-1},\frac{t}{t-7/3}} \big]$}  & {\bf see caption}                \\ \cline{3-6} 
                         &                        & $U_\tau$ & {$\mathbf{\frac{2t}{2t-3}}$}  & {\bf 27/10} ($t=3$)               & {\bf 24/7} ($t=3$)                \\ \hline
\end{tabular}
\end{table}

To prove tight bounds, we need to carefully design suitable tilings, explore the equilibrium structures, and solve related linear programming problems.

In Table~\ref{table:opt}, we list optimal assignments that will be useful for proving the results in Table \ref{tab:cylcles_cylinders_grids}. Note that these assignments are optimal for all three utility functions. In subsequent sections, we will discuss the three utility functions separately.


\begin{table}[ht] 
    \caption{Optimal assignments for $t=2$ and $t=5$ on  $C_{20}$,  $P_{40}$,  $T_{400}$. For the torus, the first 3 rows are shown for $t=2$ and the first 5 rows are shown for $t=5$. These patterns will repeat for the next 15 rows.} \label{table:opt}
    \begin{tabular}{cccccccccccccccccccccc}
        $t$ & Assignment &  \\
        2 & $L_{opt}^{C}$ & 1 & 2 & 1 & 2 & 1 & 2 & 1 & 2 & 1 & 2 & 1 & 2 & 1 & 2 & 1 & 2 & 1 & 2 & 1 & 2 \\ 
        &&&&&&&&&&&&&&&&&&&&\\  
        5  & $A_{opt}^{C}$: & 1 & 2 & 3 & 4 & 5 & 1 & 2 & 3 & 4 & 5 & 1 & 2 & 3 & 4 & 5 & 1 & 2 & 3 & 4 & 5 \\
        &&&&&&&&&&&&&&&&&&&&\\  
        2 & $L_{opt}^{P}$:   & 1 & 2 & 1 & 2 & 1 & 2 & 1 & 2 & 1 & 2 & 1 & 2 & 1 & 2 & 1 & 2 & 1 & 2 & 1 & 2 \\
        & & 2 & 1 & 2 & 1 & 2 & 1 & 2 & 1 & 2 & 1 & 2 & 1 & 2 & 1 & 2 & 1 & 2 & 1 & 2 & 1\\         
        & &&&&&&&&&&&&&           \\   
        5 & $A_{opt}^{P}$:   & 1 & 2 & 3 & 4 & 5 & 1 & 2 & 3 & 4 & 5 & 1 & 2 & 3 & 4 & 5 & 1 & 2 & 3 & 4 & 5 \\
          &           & 3 & 4 & 5 & 1 & 2 & 3 & 4 & 5 & 1 & 2 & 3 & 4 & 5 & 1 & 2 & 3 & 4 & 5 & 1 & 2\\
        &&&&&&&&&&&&&&&&&&&&&\\  
      2 &   $L_{opt}^{T}$: & 1 & 2 & 1 & 2 & 1 & 2 & 1 & 2 & 1 & 2 & 1 & 2 & 1 & 2 & 1 & 2 & 1 & 2 & 1 & 2 \\
        & & 2 & 1 & 2 & 1 & 2 & 1 & 2 & 1 & 2 & 1 & 2 & 1 & 2 & 1 & 2 & 1 & 2 & 1 & 2 & 1\\ 
         & & 1 & 2 & 1 & 2 & 1 & 2 & 1 & 2 & 1 & 2 & 1 & 2 & 1 & 2 & 1 & 2 & 1 & 2 & 1 & 2 \\    
        &          &&&&&&&&&&&&&&&&&&&&\\  
      5 &   $A_{opt}^{T}$:& 1 & 2 & 3 & 4 & 5 & 1 & 2 & 3 & 4 & 5 & 1 & 2 & 3 & 4 & 5 & 1 & 2 & 3 & 4 & 5 \\
         &           & 3 & 4 & 5 & 1 & 2 & 3 & 4 & 5 & 1 & 2 & 3 & 4 & 5 & 1 & 2 & 3 & 4 & 5 & 1 & 2\\
         &           & 5 & 1 & 2 & 3 & 4 & 5 & 1 & 2 & 3 & 4 & 5 & 1 & 2 & 3 & 4 & 5 & 1 & 2 & 3 & 4\\
          &          & 2 & 3 & 4 & 5 & 1 & 2 & 3 & 4 & 5 & 1 & 2 & 3 & 4 & 5 & 1 & 2 & 3 & 4 & 5  & 1\\
           &         & 4 & 5 & 1 & 2 & 3 & 4 & 5 & 1 & 2 & 3 & 4 & 5 & 1 & 2 & 3 & 4 & 5 & 1 & 2 & 1\\        &&&&&&&&&&&&&&&&&&&&&\\  

\end{tabular} 
\end{table}

\subsection{The binary utility function $U_b$}

In this section, we prove tight bounds on the PoA for $\sw$ and $\ce$ under $U_b$ for cycles, cylinders, and tori. In Table~\ref{table:Ub}, we list some 
worst-case assignments for these graphs, which we use to prove lower bounds on the PoA. 

\begin{table}[ht] 
  \caption{Worst-case assignments for $U_b$ for $t=2$ and $t=5$. For the torus we only show the first few rows. } \label{table:Ub}
  \begin{tabular}{cccccccccccccccccccccccccccc}
        $t$ & Assignment &  \\
        2 &$L_{b}^{C}$ & 1 & 1 & 2 & 1 & 1 & 2 & 1 & 1 & 2 & 1 & 1 & 2 & 1 & 1 & 2 & 2 & 2 & 2 & 2 & 2 \\ 
        &&&&&&&&&&&&&&&&&&&&\\  
  2 & $L_b^{P}$ & 2 & 1 & 1 & 1 & 2 & 1 & 1 & 1 & 2 & 1 & 1 & 1 & 2 & 2 & 2 & 2 & 2 & 2 &\\
               & &  1 & 1 & 2 & 1 & 1 & 1 & 2 & 1 & 1 & 1 & 2 & 1 & 2 & 2 & 2 & 2 & 2 & 2 & \\
&&&&& \\ 
   2 & $L_{b}^{T}$:   & 1 & 1 & 1 & 1 & 2 & 1 & 1 & 1 & 1 & 2 & $\cdot$ & $\cdot$ & $\cdot$ & 2 & 2 & 2 & 2 & 2 & 2 &  \\
                      & & 1 & 2 & 1 & 1 & 1 & 1 & 2 & 1 & 1 & 1 & $\cdot$ & $\cdot$ & $\cdot$ & 2 & 2 & 2 & 2 & 2 & 2 & \\
     &   & 1 & 1 & 1 & 2 & 1 & 1 & 1 & 1 & 2 & 1 &  $\cdot$ & $\cdot$ & $\cdot$ & 2 & 2 & 2 & 2 & 2 & 2 &   \\         
     &   & 2 & 1 & 1 & 1 & 1 & 2 & 1 & 1 & 1 & 1 & $\cdot$ & $\cdot$ & $\cdot$ & 2 & 2 & 2 & 2 & 2 & 2 &         \\
     & & 1 & 1 & 2 & 1 & 1 & 1 & 1 & 2 & 1  & 1 & $\cdot$ & $\cdot$ & $\cdot$ & 2 & 2 & 2 & 2 & 2 & 2 &  \\
     &  & 1 & 1 & 1 & 1 & 2 & 1 & 1 & 1 & 1 & 2 & $\cdot$ & $\cdot$ & $\cdot$ & 2 & 2 & 2 & 2 & 2 & 2 & \\
       & &&&&&&&&&&&&& \\
        5  & $A_{b}^{C}$: & 1 & 1 & 2 & 2 & 3 & 3 & 4 & 4 & 1 & 1 & 2 & 2 & 3 & 3 & 4 & 4 & 5 & 5 & 5 & 5 \\
        &&&&&&&&&&&&&&&&&&&&\\  
5  & $A_{b}^{P}$: & 1 & 1 & 2 & 2 & 3 & 3 & 4 & 4 & 1 & 1 & 2 & 2 & 3 & 3 & 4 & 4 & 5 & 5 & 5 & 5 \\
& & 1 & 1 & 2 & 2 & 3 & 3 & 4 & 4 & 1 & 1 & 2 & 2 & 3 & 3 & 4 & 4 & 5 & 5 & 5 & 5 \\
& &&&&&&&&&&&&&           \\   
5  & $A_{b}^{T}$: & 1 & 1 & 2 & 2 & 3 & 3 & 4 & 4 & 1 & 1 & 2 & 2 & 3 & 3 & 4 & 4 & 5 & 5 & 5 & 5 \\
& & 1 & 1 & 2 & 2 & 3 & 3 & 4 & 4 & 1 & 1 & 2 & 2 & 3 & 3 & 4 & 4 & 5 & 5 & 5 & 5 \\ 
& & 1 & 1 & 2 & 2 & 3 & 3 & 4 & 4 & 1 & 1 & 2 & 2 & 3 & 3 & 4 & 4 & 5 & 5 & 5 & 5 \\  
& & 1 & 1 & 2 & 2 & 3 & 3 & 4 & 4 & 1 & 1 & 2 & 2 & 3 & 3 & 4 & 4 & 5 & 5 & 5 & 5 \\ 
& &&&&&&&&&&&&&           \\   
\end{tabular}    
   \end{table}

\begin{theorem}
\label{thm:cycle-U_b}
    Consider the swap game under $U_b$ on the cycle $C_n$ with $tk$ equitable agents belonging to $t$ types. 
    \begin{enumerate}[(a)]
    \item For $t=2$, we have  $\poa(\sw,U_b, C_n) \eqas 4/3$; and $\poa(\ce,U_b,C_n) \eqas 2$;
   \item For $t \geq 3$, we have $\poa(\sw,U_b, C_n) \eqas \frac{t}{t-1}$; and $\poa(\ce,U_b,C_n) \eqas \frac{2 t}{t-1}$. 
    \end{enumerate}   
\end{theorem}

\begin{proof} The upper bounds on price of anarchy all follow from results in Section \ref{sec:Efficiency}.
We show that bounds are tight by giving  assignments that achieve these bounds. 
\begin{enumerate}[(a)] 
   
 \item  Consider the assignment $L_{opt}^{C}$ for $t=2$ in Table~\ref{table:opt}. 
 Observe that it  satisfies 
        $\sw(L_1, U_b) = 2k$
    and 
        $\ce(L_1) = 2k$.
    Next consider $L_b^{C}$ in which the type sequence '112' is repeated $k/2$ times, followed by $k/2$ many 2's; see Table~\ref{table:Ub} for an example with $k=10$. Since all type 1 agents have utility 1, they will not swap, so $L_b^{C}$ is also an  equilibrium, and all type 1 agents have utility $1$, while only $k/2 + 2 = k/2 + O(1)$ type 2 agents\footnote{The plus two comes from the two ``border'' type 2 agents. From now on, we often use $O(1)$ or $o(k)$ to indicate the count of a lower order term.} have utility $1$ and the rest have utility $0$. Hence, 
        $\sw(L_b^{C}, U_b) = k+k/2 +O(1)= 3k/2 + O(1)$,
    implying
        $\poa(\sw,U_b,C_n) \geqas 4/3$. 
    Similarly, one can verify that 
        $\ce(L_b^{C}) = k+O(1)$,
    implying
        $\poa(\ce,U_b,C_n) \geqas 2$.

    \item Consider the  assignment $A_{opt}^{C}$ in Table~\ref{table:opt}. It is easy to see that 
      $\sw(A_{opt}^{C}, U_b) = tk$
    and $\ce(A_{opt}^{C}) = tk$.  Next consider assignment $A_b^{C}$  in which the type sequence $1122\cdots (t-1)(t-1)$' is repeated $k/2$ times, followed by $t$'s repeated $k$ times; See Table~\ref{table:Ub} for an example with $k=4$.   
 It is easy to verify  that $A_b^{C}$ is at equilibrium for $U_b$, and
        $\sw(A_b^{C}, U_b) = (t-1)k + O(1)$
    and
        $\ce(A_b^{C}) = (t-1)k/2 + O(1)$.
    Hence,
        $\poa(\sw,U_b,C_n) \geqas t/(t-1)$
    and $\poa(\ce,U_b,C_n) \geqas 2t/(t-1)$. \qedhere

\end{enumerate}
\end{proof}

\begin{theorem}\label{thm:3reg-U_b}
    Consider the swap game under $U_b$ on  the 3-regular cylinder graph $P_n$ with $tk$ equitable agents belonging to $t$ types.  

    \begin{enumerate}[(a)]
        \item For $t=2$, $\poa(\sw,U_b,P_n) \eqas 3/2$ and  $\poa(\ce,U_b,P_n) \eqas 3$; 
        
        \item For $t \geq 3$, $\poa(\sw,U_b,P_n) \eqas t/(t-1)$ and $\poa(\ce,U_b,P_n) \eqas 3t/(t-1)$. 
   
 \end{enumerate}
 \end{theorem}

\begin{proof}
The upper bounds on price of anarchy all follow from results in Section \ref{sec:Efficiency}.
We show that the bounds are tight by giving specific assignments on $P_n$. . 
\begin{enumerate}[(a)]

    \item    Consider assignment $L_{opt}^{P}$ for $t=2$ from Table~\ref{table:opt}: the top row is the $L_{opt}^{C}$, the bottom row is the same but shifted by one position. It is easy to see that
        $\sw(L_{opt}^P, U_b) = 2k$
    and
        $\ce(L_{opt}^P) = 3k$.
    Now assume $k$ is a multiple of 6, and consider assignment $L_b^{P}$ for $t=2$, shown in Table~\ref{table:Ub} (for $k=18$), in which two-thirds of the agents of type 2 are segregated.  Since all agents of type 1 have utility 1, they will not swap, and  assignment  $L_b^{P}$ is an equilibrium for $U_b$. Notice that  a third of red agents also have utility $1$. Hence, 
        $\sw(L_b^P,U_b) = 4k/3 + O(1)$,
    implying
        $\poa(\sw,U_b,P_n) \geq 2k/(4k/3 +O(1)) \eqas 3/2$.
    For colorful edges, in assignment $L_b^{P_n}$, all the  agents of type 1 have exactly one colorful edge, hence,
        $\ce(L_b^P) = k + O(1)$,
    implying
        $\poa(\ce,U_b,P_n) \geqas 3$.

\item  Consider assignment $A_{opt}^{P}$  from Table~\ref{table:opt}. Observe that  
        $\sw(A_{opt}^{P}, U_b) = tk$
    and
        $\ce(A_{opt}^{P}) = 3tk/2$. Next consider  $A_b^{P}$ which is constructed by using $A_b^{C}$ in both rows of the cylinder. Observe that it is an equilibrium for $U_b$ and since there is only one type whose agents are segregated, we have
        $\sw(A_b^{P}, U_b) = (t-1)k +O(1)$
    and
        $\ce(A_b^{P}) = (t-1)k/2 +O(1)$.
    These together imply the desired lower bounds on PoA. \qedhere

\end{enumerate}

\end{proof}

\begin{theorem}\label{thm:4reg-U_b}
    Consider the swap game under $U_b$ on  the 4-regular torus $T_n$ with $tk$ equitable agents belonging to $t$ types.  

    \begin{enumerate}[(a)]
        \item For $t=2$: $\poa(\sw,U_b,T_n) \eqas 8/5$ and  $\poa(\ce,U_b,T_n) \eqas 4$;   
        
    \item For $t \geq 3$: $\poa(\sw,U_b,T_n) \eqas t/(t-1)$ and $\poa(\ce,U_b,T_n) \eqas 4t/(t-1)$.  
\end{enumerate}
 \end{theorem}

\begin{proof}
The upper bounds on price of anarchy all follow from results in Section \ref{sec:Efficiency}.
We give assignments that meet these bounds. 
    \begin{enumerate}[(a)]

      \item   Consider the assignment $L_{opt}^T$ shown in Table~\ref{table:opt}. It is easy to observe that $\sw(L_{opt}^{T}, U_b) = tk$ and
        $\ce(L_{opt}^{T}) = 4k$. Next assume $k=32i^2$ where $5 \mid i$ and consider assignment $L_b^T$ shown in Table~\ref{table:Ub}. It is constructed as follows: In the first row, repeat $i$ times the type sequence $11112\cdots$, and follow by $3i$ times '2';   in the second row, repeat $i$ times the sequence $12111\cdots$, third row: $11121$, fourth row: $21111$, fifth row: $11211$, and in each row follow by  $3i$ times 2. Repeat these 5 rows until all agents are assigned.  
          This can be verified to be an equilibrium in which  except for 'border' agents, all agents of type 1, and $1/4$ of type  $2$ agents have utility $1$, so that $\sw(L_b^{T}, U_b) \leq k+k/4 + o(k)$, hence $\poa(\sw,U_b,T_n) \geqas 2k/(k+k/4) = 8/5$. For colorful edges,  all the non-boundary agents of type 1 have exactly one colorful edge, so that $\ce(L_b^{T}) \leq k + o(k)$ This implies that  $\poa(\ce,U_b,T_n) \geqas 4$.

        \item Consider the assignment $A_{opt}^T$ shown in Table~\ref{table:opt}. It is easy to observe that $\sw(A_{opt}^{T}, U_b) = tk$ and
        $\ce(A_{opt}^{T}) = 2tk$. Next assume $k=4ti^2$ and consider  $L_b^{T}$ shown in Table~\ref{table:Ub}, which is constructed by using $L_b^{C}$ in every row of the torus. Observe that it is an equilibrium for $U_b$ and since  all but $o(k)$ agents of type $t$ are segregated, and agents of every other type as well as $o(k)$ agents of type $t$ have exactly one colorful edge, we have
        $\sw(A_b^{T}, U_b) = (t-1)k + o(k)$
    and
        $\ce(A_b^{T}) = (t-1)k/2 +o(k)$.
    These together imply the desired lower bounds of PoA.       \qedhere 

    \end{enumerate}
\end{proof}

\subsection{The variety-seeking utility function $U_\tau$}

In this section, we prove tight bounds on the PoA for $\sw$ and $\ce$ under $U_\tau$ for cycles, cylinders, and tori. In Table~\ref{table:U-tau}, we list some 
worst-case assignments for these graphs, which we use to prove lower bounds on the PoA. 

\begin{table}[!htbp] 
\caption{Worst-case assignments for $U_\tau$ for $t \geq 3$; for $t=2$, refer to $L_b^C$, $L_b^P$, and $L_b^T$ from Table~\ref{table:Ub}. Assignments $A_\tau^C$, $A_\tau^P$, and $A_\tau^T$ minimize the social welfare, while $B_\tau^C$, $B_\tau^P$, and $B_\tau^T$ minimize the number of colorful edges. For the torus we only show a partial assignment. }\label{table:U-tau}
 \resizebox{\textwidth}{!}{%
 \begin{tabular}{cccccccccccccccccccccccccccccccccccc}
     Assignment &  \\
     $A_{\tau}^{C}$: & 1 & 2 & 1 & 2 & 1 & 2 & 1 & 2 & 3 & 4 & 3 & 4 & 3 & 4 & 3 & 4 & 5 & 5 & 5 & 5 \\
        &&&&&&&&&&&&&&&&&&&&\\  
        $B_\tau^C$ & 1 & 2 & 1 & 1 & 2 & 1 & 1 & 2 & 1 & 2 & 3 & 2 & 2 & 3 & 2 & 1 & 2 & 1 &  3 & 3 & 3 & 3 & 3 & 3 \\
             &&&&&&&&&&&&&&&&&&&&\\  
  $A_{\tau}^{P}$: & 1 & 2 & 1 & 2 & 1 & 2 & 1 & 2 & 3 & 4 & 3 & 4 & 3 & 4 & 3 & 4 & 5 & 5 & 5 & 5 \\
& 2 & 1 & 2 & 1 & 2 & 1 & 2 & 1 & 4 & 3 & 4 & 3 & 4 & 3 & 4 & 3 & 5 & 5 & 5 & 5 \\
& &&&&&&&&&&&&&           \\   
$B_{\tau}^{P}$: & \mg{3} & \mg{3} & \mg{3} & \mb{1} & $\cdot$ & 1 & 2 & 1 & \mb{1} & \mr{2} & \mr{3} & \mr{2} & 2 & $\cdot$ & \mr{2} & \mr{3}  & \mr{2} & 1 & 3 & 3 & 3 & \mg{3} & $\cdot$ \\
&  & \mg{3} & \mb{1} & \mb{2} & \mb{1} & $\cdot$  & 1 & \mb{1} & \mb{2} & \mb{1} & \mr{2} & 2 & 3 & 2 & $\cdot$ & \mr{2} & 1  & 2 & 1 & 3 & \mg{3} & \mg{3} & \mg{3} \\
& &&&&&&&&&&&&&           \\   
 $A_{\tau}^{T}$: & 1 & 2 & 1 & 2 & 1 & 2 & 1 & 2 & 3 & 4 & 3 & 4 & 3 & 4 & 3 & 4 & 5 & 5 & 5 & 5 \\
 & 2 & 1 & 2 & 1 & 2 & 1 & 2 & 1 & 4 & 3 & 4 & 3 & 4 & 3 & 4 & 3 & 5 & 5 & 5 & 5 \\
& 1 & 2 & 1 & 2 & 1 & 2 & 1 & 2 & 3 & 4 & 3 & 4 & 3 & 4 & 3 & 4 & 5 & 5 & 5 & 5 \\
& 2 & 1 & 2 & 1 & 2 & 1 & 2 & 1 & 4 & 3 & 4 & 3 & 4 & 3 & 4 & 3 & 5 & 5 & 5 & 5 \\
& &&&&&&&&&&&&&           \\   
$B_\tau^T$ & &&&&&&&&&\bd&\bd&\bd&&&&&&&&&\\
& &&&&\mg{3}&&&&&3&&&&&3&&&&&&\\ 
& &&\mb{1}&\mg{3}&\mg{3}&\mg{3}&&\mb{1}&3&3&3&&1&3&3&3&&1&3&3&3\\ 
& &\mb{1}&\mb{2}&\mb{1}&\mg{3}&1&\mb{1}&\mb{2}&\mb{1}&3&1&1&2&1&3&1&1&2&1&3&1\\
& &&\mb{1}&1&1&2&1&\mb{1}&1&1&2&1&1&1&1&2&1&1&1&1&2&1\\
& &&1&2&1&1&1&1&2&1&1&\mb{1}&1&2&1&1&1&1&2&1&1&1\\
& &&&1&2&1&2&1&1&2&\mb{1}&\mb{2}&\mb{1}&1&2&1&2&1&1&2&1&1&1\\ 
&.&&\mr{2}&2&3&2&1&\mr{2}&2&3&2&\mb{1}&2&2&3&2&1&2&2&3&1&2\\ 
&.&\mr{2}&\mr{3}&\mr{2}&2&2&\mr{2}&\mr{3}&\mr{2}&2&2&2&3&2&2&2&2&3&2&2&2\\
&.&&\mr{2}&&2&3&2&\mr{2}&&2&3&2&2&&2&3&2&2&&2&3&2\\
&&&&&&2&&&&&2&&&&&2&&&&&2\\
& &&&&&&&&&\bd&\bd&\bd&&&&&&&&&\\ 
& &&&\mr{2}&&&&&2&&&&&2&&&&&2&&\\ 
&&&\mr{2}&\mr{3}&\mr{2}&&1&2&3&2&&1&2&3&2&&1&2&3&2&&1\\  
& &&&\mr{2}&1&1&2&1&2&1&1&2&1&2&1&1&2&1&2&1&1&2&1\\ 
& &&1&1&2&1&1&1&1&2&1&1&1&1&2&1&1&1&1&2&1&1\\ 
&&1&2&1&1&3&1&2&1&1&3&1&2&1&1&3&1&2&1&1&3\\
&&&1&\mg{3}&3&3&3&1&\mg{3}&3&3&3&1&3&3&3&3&1&3&3&3&3\\
&&&\mg{3}&\mg{3}&\mg{3}&3&3&\mg{3}&\mg{3}&\mg{3}&3&3&3&3&3&3&3&3&3&3&3&3\\
&&&&\mg{3}&&3&3&3&\mg{3}&&3&3&3&3&&3&3&3&3&&3&3&3\\
&&&&&&&3&&&&&3&&&&&3&&&&&\\ 
&&&&&&&&&&&\bd&\bd&\bd&&&&&&&&&\\ 
\\
& &&&&&&&&&&&&&           \\   
\end{tabular} }
 \end{table}

\begin{theorem}\label{thm:cycle-U_tau}
    Consider the swap game under $U_\tau$  on the cycle $C_n$ with $tk$ equitable agents belonging to $t\ge 3$ types.
    \begin{enumerate}[(a)]
\item  $\poa(\sw, U_\tau, C_n)   \eqas \frac{2t}{t-1}$; 
\item $\poa(\ce, U_\tau, C_n) \eqas \frac{2t}{2t-3}$.  

  \end{enumerate}
\end{theorem}

\begin{proof}
\begin{enumerate}[(a)]
\item Consider assignment $A_{opt}^C$ from Table~\ref{table:opt}. It is easy to check that \\ $\sw(A_{opt}^C, U_\tau) = 2tk$. By Observation \ref{obs:utility_observation}, at most one type say $t$ can have segregated agents, and  agents of other types all have utility at least $1$. This means the social welfare for any equilibrium assignment is at least $(t-1)k$. For odd $t$, this is achieved by assignment $A_\tau^C$ shown in Table~\ref{table:U-tau},  in which agents of type $t$ are segregated, and agents of remaining types are paired into groups of two, and agents in each pair alternate with each other.  It can be verified that $A_\tau^C$ is an equilibrium, in which all but 2 agents of type $t$ have utility 0, and except for $O(t)$ 'border' agents, all other agents of all other types have only one type in their neighborhood. Therefore the social welfare is $(t-1)k + O(t)$, implying the tight bound of $\frac{2t}{(t-1)}$ on $\poa(\sw,U_\tau,C_n)$. 
For even $t$, we can give a similar construction: for the first 3 types, we repeat $k/2$ times the type sequence $12$, then we repeat $k/2$ times the sequence $23$ followed by $k/2$ repetitions of $13$. The remaining types other than type $t$ are paired, and we have alternating sequences of length $2k$ with the types in a pair. Finally we have a sequence of agents of type $t$.  For example, for $t=6$, and $k=6$, we can assign agents according to the type sequence
$121212131313232323454545454545666666$.

\item Next we consider the number of colorful edges. In Assignment $A_{opt}^C$, all $tk$ edges are colorful. We now show that in every equilibrium there are at most $3k/2$ monochromatic edges.  By Corollary~\ref{cor:U_tau_num_of_types}, there are at most three types whose agents can have one colorful edge and one monochromatic edge. 

Case i: If  there are no segregated agents, then there are at most $3k$ agents who are incident on a monochromatic edge, and since none of them is segregated, there are at most $3k/2$ monochromatic edges as claimed. 

Case ii: Now we consider the case when there are segregated agents. Assume wlog that there exist agents of type $t$ that are segregated.  By Observation \ref{obs:utility_observation}, agents of all other types are not segregated. There must be an agent $A$ of type $t$ that has a neighbor of its own type as well as agent of another type, wlog, let $T(A) = \{t, 1 \}$. Consider now an agent $B$ of any other type, say $x\notin \{1, t\}$ that has a monochromatic edge with an agent $C$ also of type $x$.  If $T(B) = \{x, y \}$ for $y \neq t$, then $A$ and $B$ would swap, contradicting the assumption of an equilibrium. Therefore, it must be that $T(B) = \{x, t\}$. By the same reasoning, we have $T(C)= \{x, t \}$. Thus, the sequence of types in $B$ and $C$'s neighborhood is $txxt$. 
To summarize, we have shown that there are only three different cases of monochromatic edges: (1) edge $tt$,  (2) edge $11$, and (3) edge $xx$ with $x \neq 1, t$, and edge $xx$ only exists in the sequence $txxt$. For monochromatic edge $tt$ we say the right $t$ is \emph{responsible} for this edge, and for monochromatic edge $xx$ we say the right $t$ in $txxt$ is responsible. Then, every monochromatic $tt$ or $xx$  edge has a unique type $t$ agent that is responsible. This shows the number of such monochromatic edges is at most the number of type $t$ agents, i.e., at most $k$. Because type $1$ agents are not segregated, the number of $11$ edges is at most $k/2$. Hence, the total number of monochromatic edges  is at most $3k/2$ as claimed. Since the number of colorful edges is then at least $t-3k/2$, we obtain $\poa(\ce, U_\tau, C_n) \leq \frac{tk}{t - 3k/2} = \frac{2t}{t-3}$.
    Finally, the lower bound on the PoA is achieved by assignment $B_\tau^C$ for $t=3$, which consists of $k/2-1$ repetitions of the type sequence $121$ followed by $k/4$ repetitions of $232$, then the sequence $121$ followed by $3k/4$ type 3 agents. $B_\tau^C$ can be verified to be an equilibrium, as the swap condition of Lemma~\ref{lem:swap-condition-U_tau} is not met, and it has $3k/2$ colorful edges.  For $t>3$, it is possible to achieve the bound with a similar construction where type $t$ agents are segregated in one 'block', the border agents of this block have neighbors of type 1, which is the only other type to have monochromatic edges. For example, for $t=5$, the assignment  $121|131|234|234|234|234|141|555555$ achieves the lower bound. \qedhere

    \end{enumerate}
    
    \end{proof}

\begin{theorem} \label{thm:3reg-U_tau}
   Consider a swap game under $U_\tau$ on the 3-regular cylinder graph $P_n$ with a set of $tk$ equitable agents belonging to $t\geq 3$ types.
   \begin{enumerate}[(a)]
        \item $\poa(\sw, U_\tau, P_n) \eqas \min \{t, \frac{3t}{t-1} \}$; 
        \item For $t = 3$, we have $\poa(\ce, U_\tau, P_n) \eqas 27/10$.

    \end{enumerate}
\end{theorem}

\begin{proof}

(a)   
The upper bound on $\poa(\sw, U_\tau, P_n)$ follows from Section \ref{sec:Efficiency}. 
Here we show that the bound is tight for $P_n$. Consider assignment $A_{opt}^P$ from Table~\ref{table:opt}. It is easy to check that  $\sw{A_{opt}^P, U_\tau} = \min \{ t-1, 3 \} tk$.  By Observation \ref{obs:utility_observation}, at most one type say $t$ can have segregated agents, and  agents of other types all have utility at least $1$. This means the social welfare for any equilibrium assignment is at least $(t-1)k$. For odd $t$, this is achieved by assignment $A_\tau^P$ shown in Table~\ref{table:U-tau},  in which 
we place the assignment $A_\tau^C$ for the cycle (from Table~\ref{table:U-tau}) in the first row, and the same in the second row except that the agents of type $i$ and $i+1$ exchange places. It can be verified that $A_\tau^P$ is an equilibrium, in which all but 2 agents of type $t$ have utility 0, and except for $O(t)$ 'border' agents, all other agents of all other types have only one type in their neighborhood. Therefore the social welfare is $(t-1)k + O(t)$, implying that $\poa{sw}{U_\tau} \geq \min \{t, \frac{3t}{t-1} \}$ as desired.  
A similar assignment can also be given for $t$ even.

    (b)  Consider assignment $A_{opt}^P$ in Table~\ref{table:opt}; it is easy to see that $\ce(A_{opt}^P) = 9k/2$.  
    We show that every equilibrium must have at least $5k/3$ colorful edges, and this can be (asymptotically) achieved 
    by assignment $B_\tau^P$ in Table~\ref{table:U-tau}.
    Hence, the PoA is
        $(9k/2)/(5k/3) = 27/10$.
        
    Firstly, consider $B_\tau^P$ created by tiling the cylinder with 3 classes of tiles:  $R$ (type-3 surrounded by type-2), 
        $B$ (type-2 surrounded by type-1), and $G$ (type-3 surrounded by type-3). One can verify that it is at equilibrium under $U_\tau$ as only type-3 agents have their own type in their neighborhood, and so the swap condition in Lemma~\ref{lem:swap-condition-U_tau}  is not met. Let the number of tiles of class $R$, $B$, and $G$ is respectively, $r$, $b$, $g$. Counting the number of $1$'s, $2$'s, and $3$'s, we have,
    \begin{equation}
        \begin{cases}
            3b &= k; \\
            b + 3r &= k; \\
            r + 4g &= k.
        \end{cases}
    \end{equation}
    Solving the above, we get
        $r = 2k/9, b =k/3, g = 7k/36$.
    Note that each  $R$ tile contains three colorful edges, and each $B$ tile contains three colorful edges, and there are no other colorful edges except $O(1)$ many colorful edges in the boundary connecting different classes of tiles. Hence, the total number of colorful edges is 
        $3(r+b) + O(1) = 5k/3 + O(1)$. 

   Next, we show that every equilibrium must have at least $5k/3$ colorful edges. By Observation \ref{obs:utility_observation},  we assume without loss of generality that only color $3$ can possibly be segregated, i.e.,  every agent of type $1$  or $2$ has utility at least one. Consider two cases.

    (i)
        Every agent of type $3$ has at least two colorful edges, hence the total number of colorful edges is at least $2k$.

        (ii) There exists some agent of type $3$ that has at most one colorful edge. Clearly, it is impossible for all agents of type $3$ to be segregated. Hence, there exists some type 3 agent call it $A$, located at say $v$, that has exactly one colorful edge. WLOG, assume its neighborhood is of the form $\{1,3,3\}$. 

        {\bf Claim.} Every type 2 agent satisfies the following: 
            either its neighborhood is exactly of the form $\{1,1,1\}$,
        or its neighborhood contains a color $3$. 

        Proof of Claim:  Consider an arbitrary type 2 agent $B$ located at vertex $v'$, and consider the colors in $N(v')$. Assume for the sake of a contradiction that $N(v')$ does not satisfy the Claim, then, $N(v')$ contains both\footnote{Note, here we used the fact that color $2$ can not be segregated, so, $\{2,2,2\}$ is not an option.} color $1$ and $2$ but not $3$, then, $A$ and $B$ will swap and both increase their utilities from one to two, a contradiction to the equilibrium.

        Now, let $x$ denote the number of type 2 agents whose neighborhood is of the form $\{1,1,1\}$, and $y$ denote the rest type 2 agents. Then, 
        \begin{equation}    \label{eq:U_t_3_LP1-another}
            x + y = k.
        \end{equation}
        Furthermore, let $y_i$ denote the number of agents of the second category  whose neighborhood contains exactly $i$  agents of color $1$ (besides an agent of color $3$ guaranteed by Claim), for $i=0,1,2$. Then,
        \begin{equation}    \label{eq:U_t_3_LP2-another}
            y_0 + y_1 + y_2 = y.
        \end{equation}
        Let $p$ denote the number of type $1$ agents that has at least one type $2$ agent in its neighborhood, and $q$ denote the number of type $1$ agents that has at least one type $3$ agent in its neighborhood. Trivially,
        \begin{equation}    \label{eq:U_t_3_LP3-another}
            p+q \ge k.
        \end{equation}
        Now, we count the number of colorful edges. Let
            $c_{ij}$
        denote the number of colorful edges with color $i$ and $j$ in its two ends. Then, 
        \begin{equation}    \label{eq:U_t_3_LP4-another}
            \begin{cases}
            c_{12} &= 3x+ 2y_2 + y_1  \\
            c_{23} &\ge y  \\
            c_{13} &\ge q.
            \end{cases}
        \end{equation}
        Note also that 
        \begin{equation}    \label{eq:U_t_3_LP5-another}
            p \le c_{12}.
        \end{equation}
        Now, to prove a lower bound for the number of colorful edges is equivalent to minimize the objective function
            $c_{12} + c_{23} + c_{13}$
        under the linear constraints \eqref{eq:U_t_3_LP1-another}-\eqref{eq:U_t_3_LP5-another}.
        Since both the objective function and the constraints are linear, we can replace constraint \eqref{eq:U_t_3_LP1-another} by $x+y=1$ and  constraint \eqref{eq:U_t_3_LP1-another} by $p+q \ge 1$, solving this linear programming gives $OPT = 5/3$, which implies the original LP OPT is $5k/3$. 
    In all cases, the number of colorful edges is at least $5k/3$ as claimed. \qedhere
\end{proof}

\begin{theorem}\label{thm:4reg-U_tau}
   Consider a swap game under $U_\tau$ on the 4-regular torus graph $T_n$ with a set of $tk$ equitable agents belonging to $t\geq 3$ types.
    \begin{enumerate}[(a)]

        \item $\poa(\sw, U_\tau, T_n) \eqas  \min\{t, \frac{4 t}{t-1} \}$;  
        \item For $t = 3$, we have $\poa(\ce, U_\tau, T_n) \eqas 24/7$.    
    \end{enumerate}    
\end{theorem}

\begin{proof}
    \begin{enumerate}[(a)]
    \item 
The upper bound follows from Section \ref{sec:Efficiency}. 
Here we show that the bound is tight for $T_n$. Consider assignment $A_{opt}^T$ from Table~\ref{table:opt}. It is easy to check that  $\sw{A_{opt}^P, U_\tau} = \min \{ t-1, 4 \} tk$.  By Observation \ref{obs:utility_observation}, at most one type say $t$ can have segregated agents, and  agents of other types all have utility at least $1$. This means the social welfare for any equilibrium assignment is at least $(t-1)k$. For odd $t$, this is achieved by assignment $A_\tau^T$ shown in Table~\ref{table:U-tau},in which we repeat the assignment $A_\tau^P$ for the cylinder until all rows are assigned.  This ensures that all except ``border'' agents of all types $\neq t$  have only one other type in their neighborhoods, and furthermore, no agent except the agents of type $t$ ever have their own type in their neighborhood. This ensures that the swap condition in Lemma~\ref{lem:swap-condition-U_tau} is not met and we have an equilibrium assignment. T
Therefore the social welfare is $(t-1)k + O(t)$, implying that $\poa(\sw,U_\tau,T_n) \geq \min\{t, \frac{4 t}{t-1} \}$ as desired.  
A similar assignment can also be given for $t$ even. 

        \item  The lower bound: the assignment in Figure \ref{fig:opt_diversity_example}-(b) satisfies all $(3k) \cdot 4/2 = 6k$ edges are colorful. 
        Consider now $B^T_\tau$ in Table~\ref{table:U-sharp}: it is an equilibrium by Lemma \ref{lem:swap-condition-U_tau}, because only type-3 agents have their own type in their neighborhood. Note that there are three classes of tiles: $R$ (type-3 surrounded by type-2), $B$ (type-2 surrounded by type-1), and $G$ (type-3 surrounded by type-3). Suppose the number of $R$, $B$, and $G$ tiles is $r, b, g$ respectively. Counting the number of type-1, type-2 and type-3 agents, we have
        $4b = k$, 
        $b + 4r = k$, and
        $r + 5g = k$.
    Solve it we get
        $r = 3k/16, b =k/4, g = 13k/80$.
    Note that each $R$ and $B$ contains four colorful edges, and there are no other colorful edges except $o(k)$ many colorful edges in the boundary connecting different classes of tiles. Hence, 
        $\ce(L^e) = 4(r+b) + o(k) \lesssim 7k/4$,
    as desired.

    The upper bound: it suffices to show $\ce(L) \ge 7k/4$ for every equilibrium $L$. By Observation \ref{obs:utility_observation}-(2), wlog we assume only type-3 agents can be segregated. If every type-3 agent has at least two colorful edges, then we are done. Hence, suppose some type-3 agent $A$ has exactly one colorful edge (note that clearly it is impossible for all type-3 agents to have monochromatic neighborhood). We assume wlog that $A$'s neighbors are of the form $\{1,3,3,3\}$, i.e., one type-1 agent and three type-3 agents. Apply Lemma \ref{lem:swap-condition-U_tau}, we observe that: the neighborhood of every type-2 agent is either of the form $\{1,1,1,1\}$ or contains a type-3 agent, see Figure \ref{fig:grid_proof}.

    \begin{figure}[ht!]
        \centering
      \includegraphics[scale=.7]{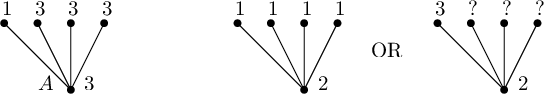}
        \caption{The neighborhood structures.} 
        \label{fig:grid_proof}
    \end{figure}    

    For type-2, let $x$ denote the number of agents whose neighborhood is of the form $\{1,1,1,1\}$, and $y$ the rest. Furthermore, let $y_i$ denote the number of agents of the second class  whose neighborhood contains exactly $i$  type-$1$ agents, for $i=0,1,2,3$. Then,
        \begin{align}   
            x + y &= k, \label{eq:U_t_3_LP1} \\
            y_0 + y_1 + y_2 + y_3 &= y.   \label{eq:U_t_3_LP2} 
        \end{align}

        Let $p$ (resp. $q$) denote the number of type-$1$ agents that has at least one type-$2$ (resp. type-$3$) agent in its neighborhood. Then,
        \begin{equation}    \label{eq:U_t_3_LP3}
            p+q \ge k.
        \end{equation}
        We now count the number of colorful edges. Let
            $c_{ij}$
        denote the number of colorful edges between type-$i$ and type-$j$ agents. Then, 
        \begin{equation}    \label{eq:U_t_3_LP4}
            \begin{cases}
            c_{12} &= 4x + 3y_3 + 2y_2 + y_1.  \\
            c_{23} &\ge y  \\
            c_{13} &\ge q.
            \end{cases}
        \end{equation}
        Note also that 
        \begin{equation}    \label{eq:U_t_3_LP5}
            p \le c_{12}.
        \end{equation}
        Now, to prove a lower bound for the number of colorful edges is equivalent to minimize the objective function
            $c_{12} + c_{23} + c_{13}$
        under the linear constraints \eqref{eq:U_t_3_LP1}-\eqref{eq:U_t_3_LP5}.
        Since both the objective function and the constraints are linear, we can replace constraint \eqref{eq:U_t_3_LP1} by $x+y=1$ and  constraint \eqref{eq:U_t_3_LP3} by $p+q \ge 1$. Solve the LP gives the OPT is $7/4$, i.e., the number of colorful edges is at least $7k/4$ as desired.\qedhere

 \end{enumerate}
\end{proof}

\subsection{The difference-seeking utility function $U_\#$}

In this section, we prove tight bounds on the PoA for $\sw$ and $\ce$ under $U_\#$ for cycles, cylinders, and tori. In Table~\ref{table:U-sharp}, we list some worst-case assignments for these graphs, which we use to prove lower bounds on the PoA; we will also use some assignments from Table~\ref{table:U-tau} for this purpose. 

\begin{table}[!htbp] 
\caption{Worst-case assignments for $U_\#$ for $t=2$. In $A_\#^T$ for the torus, the agents in bold have utility 3, while the others have utility 2.
For $t \geq 3$, refer to $A_\tau^C$, $A_\tau^P$ and $A_\tau^T$ from Table~\ref{table:U-tau}. } \label{table:U-sharp} 
\resizebox{\textwidth}{!}{%
\begin{tabular}{ccccccccccccccccccccccccccccccccc}
  \setlength{\tabcolsep}{1pt}
  $t$ & Assignment &  \\
       2  & $L_{\#}^{C}$: & 1 & 1 & 2 &  1 & 1 & 2   & \bf{1} & \bf{2} & \bf{1} & 2 & 2 & 1 & 2 & 2 & 1 & \bf{2} & \bf{1} & \bf{2}  \\
                   &&&&&&&&&&&&&&&&&&&&\\  
     2  & $L_{\#}^{P}$: & 1 & 2 & 1 &  2 & 1 & 2   & 1 & 2 & 1 & 2 & 1 & 2 & 1 & 2 & 1 & 2 & 1 & 2  \\
& & 1 & 2 & 1 &  2 & 1 & 2   & 1 & 2 & 1 & 2 & 1 & 2 & 1 & 2 & 1 & 2 & 1 & 2  \\
                  &&&&&&&&&&&&&&&&&&&&\\  

   2 & $L_\#^T$       &&&&&&&&&\bd&\bd&\bd&&&&&&&&&\\ 
& & &\bt&1&1&\bt&2&\bo&2&2&\bo&2&\bt&1&1&\bt&2&\bo&2&2&\bo&2&\\  
& & &1&\bt&1&1&\bt&1&\bo&2&2&\bo&1&\bt&1&1&\bt&1&\bo&2&2&\bo&\\ 
& & \bd &\bo&2&\bt&1&1&\bt&2&\bo&2&2&\bo&2&\bt&1&1&\bt&2&\bo&2&2&\bd\\
& & .\bd &2&\bo&1&\bt&1&1&\bt&1&\bo&2&2&\bo&1&\bt&1&1&\bt&1&\bo&2&\bd\\
& & \bd &2&2&\bo&2&\bt&1&1&\bt&2&\bo&2&2&\bo&2&\bt&1&1&\bt&2&\bo&\bd\\
& & &\bo&2&2&\bo&1&\bt&1&1&\bt&1&\bo&2&2&\bo&1&\bt&1&1&\bt&1&\\
& & &2&\bo&2&2&\bo&2&\bt&1&1&\bt&2&\bo&2&2&\bo&2&\bt&1&1&\bt&\\
& & &\bt&1&\bo&2&2&\bo&1&\bt&1&1&\bt&1&\bo&2&2&\bo&1&\bt&1&1&\\
& & &&&&&&&&&\bd&\bd&\bd&&&&&&&&&\\ 
& &&&&&&&&&&&&&           \\   
 \end{tabular}}
\end{table}

\begin{theorem}\label{thm:cycle-U_sharp}
   Consider the swap game under $U_\#$ on $C_n$ with a set of $tk$ equitable agents belonging to $t$ types. 
    \begin{enumerate}[(a)]
        \item For $t=2$,  we have $\poa(\sw,U_\#, C_n) \eqas 3/2$;
            \item For $t \geq 3$, we have $\poa(\sw,U_\#, C_n) \eqas  \frac{t}{t-1}$.        
            \end{enumerate}
\end{theorem}

\begin{proof} 

\begin{enumerate}[(a)]
    \item The assignment $L_{opt}^C$ has a social welfare of $4k$, as each of the $2k$ agents has utility 2. Consider assignment $L_\#^C$ in Table~\ref{table:U-sharp}: we have $k/3-1$ repetitions of $112$, followed by a ``border'' sequence of $121$ followed by $k/2$ repetitions of $221$ followed by $212$. It is easy to 
    verify that $L_\#^C$ is an equilibrium and $\sw(L_\#^c, U_\#) \geq 8k/3 + O(1)$, hence $\poa(\sw,U_\#, C_n) \geqas 4k/(8k/3) = 3/2$. 
    
    We now show 
            $\sw(L,U_\#) \ge 8k/3$
        for any equilibrium assignment $L$.  
        Let $L$ be an arbitrary equilibrium under $U_\#$. 
        Let $k_{1i}$ and $k_{2i}$ denote the number of type 1 and type 2 agents, respectively, of utility $i$, for $i=1,2$. By Lemma~\ref{lem:U_sharp_basic_property}, for every type 1 agent of utility one, its type 2 agent neighbor must have utility two. Since a type 2 agent can be a neighbor of at most two type 1 agents, 
            $k_{11} \le 2k_{22}$.
        A symmetrical argument shows that
            $k_{21} \le 2k_{12}$.
        Trivially,
            $k_{11} + k_{12} = k_{21} + k_{22} = k$. 
        This together with the previous two inequalities implies that
            $k_{12} + k_{22} \ge 2k/3$. 
        Hence, $\sw(L, U_\#) \geq 
            (k_{11} + k_{21}) + 2 \cdot (k_{12} + k_{22}) 
            \geq (k_{11} + k_{21} + k_{12} + k_{22}) + k_{12} + k_{22} 
            \geq 2k + 2k/3 =  8k/3$.
            This shows that $\poa(\sw,U_\#, C_n) \leq \frac{4k}{8k/3} = 3/2$.

\item The optimal social welfare is  $2tk$, given by assignment $L_{opt}^C$. Consider $A_{\tau}^C$ from Table~\ref{table:U-tau}  in which
agents of type $t$ are segregated, and agents of remaining types are paired into groups of two\footnote{Assuming $t$ is odd. If $t$ is even, a similar assignment can be given.}, and agents in each pair alternate with each other. This
is also an equilibrium for $U_\#$ as all agents except type-$t$ agents have maximum utility of 2, and $\sw(A_\tau^C, U_\#) \geq 2(t-1)k + O(1)$. This shows that $\poa(\sw,U_\#, C_n) \geqas t/(t-1)$. We now show that the bound is tight. We claim that the social welfare at any equilibrium is at least $2(t-1)k$. If there is a segregated node, then by Lemma~\ref{lem:U_sharp_basic_property}, agents of all other types must  have utility $2$, hence, the social welfare is at least $2(t-1)k$ as desired. So assume 
there is no segregated node. By Corollary~\ref{cor:U_sharp_num_of_types}, there are at most three types whose agents can have utility $1$. 

Next we show that there are at most $2k$ agents that can have utility $1$, and the remaining must have utility $2$. This is trivially true if there are at most two types that have utility $1$ since each type has at most $k$ agents. So suppose there are exactly three types, say type $x, y, z$, of agents that can have utility $1$. By Lemma~\ref{lem:swap-condition-U_tau}, we can conclude without loss of generality that every agent of type $x$ that has utility 1 has a neighbor of type $y$, every agent of type $y$ that has utility 1 has a neighbor of type $z$, and every agent of type $z$ that has utility 1 has a neighbor of type $x$. In other words, we have sequences of agent types $xxy$, $yyz$, and $zzx$. Furthermore, these sequences are non-overlapping (that is, we cannot have the pattern $xxyy$, otherwise the center two would swap). Out of every such triple, two agents have utility $1$. Thus, out of a total of $3k$ agents of these three types, at most $2k$ can have utility $1$ as claimed. It follows  that the total social welfare is at least 
    $2(t-2)k + 2k = 2(t-1)k$.  This completes the proof that $\poa(\sw,U_\#, C_n) \leq \frac{2tk}{2(t-1)k} = \frac{t}{t-1}$. 
\end{enumerate}
\end{proof}

\begin{theorem}\label{thm:3reg-U_sharp}
   Consider the swap game under $U_\#$ on $P_n$ with a set of $tk$ equitable agents belonging to $t$ types. 
    \begin{enumerate}[(a)]

      \item For $t=2$,  we have $\poa(\sw,U_\#, P_n) \eqas 3/2$ .
   
       \item For $t \geq 3$, we have $t/(t-1) \leqas \poa(\sw,U_\#, P_n) \le t/(t-7/3)$. 
  
    \end{enumerate}   
\end{theorem}

\begin{proof} 
    \begin{enumerate}[(a)]
        \item  In Assignment $L_{opt}^P$ from Table~\ref{table:opt}, every agent has utility 3, and \\$\sw(L_{opt}^P, U_\#) = 6k$. On the other hand, assignment $L_\#^P$ from Table ~\ref{table:U-sharp} is an equilibrium and $\sw(L_{opt}^P, L_\#) =4k$. Therefore $\poa(\sw,U_\#, P_n) \geq 3/2$. We show now that the social welfare of any assignment at equilibrium is at least $4k$. By Lemma~\ref{lem:U_sharp_basic_property}, no agent can be segregated, that is all agents have utility at least 1.  If all agents have utility at least 2, then we are done.   Suppose instead that there is a type-1 agent which has utility 1. Then by Lemma~\ref{lem:U_sharp_basic_property}, all type-2 agents have utility at least 2. 
        Let $r_i$ and $b_i$ be the number of type-1 and type-2 agents resp. with utility $i$, that is, with $i$ colorful edges. 
     Then, 
            $r_1 + r_2 + r_3 = k = b_2 + b_3$.
        Counting the number of colorful edges, one has
            $r_1 + 2r_2 + 3r_3 = 2b_2 + 3b_3$.
        Thus,
            $k + r_2 + 2r_3 = 2k + b_3$
        which implies
            $r_2 + 2r_3 = k + b_3 \ge k$.
        Hence, the social welfare is
            $r_1 + 2r_2 + 3r_3 + 2b_2 + 3b_3
            = (k + r_2 + 2r_3 + 2k + b_3
            = 2k+ b_3 + 2k + b_3 \ge 4k$
        as claimed.

 \item Assignment $A_{opt}^P$ from Table~\ref{table:opt} has social welfare $3tk$. On the other hand consider $A_{\tau}^P$ from Table~\ref{table:U-tau}, in which the assignment $A_\tau^C$ is in the first row, and the same in the second row except that agents of type $i$ and $i+1$ switch places for odd $i$. This is also an equilibrium for $U_\tau$ in which  every agent of types other than $t$ has utility 3 while all but two of the agents of type $t$ are segregated.  and $\sw(A_{\tau}^P, U_\#) = (t-1)k + O(1)$, hence $\poa{sw}{U_\#} \geq \frac{tk}{(t-1)k} = t/(t-1)$. We also conclude that $\wdoit{2}=0$, as except for the 'border' agents, all other agents of type $\neq t$ have agents of exactly one type as neighbors.

  We will now show that  $\sw(L, U_\#) \ge (3t-7)k$ for any equilibrium assignment $L$.
  
    By Corollary~\ref{cor:U_sharp_num_of_types}, the number of types whose agents can have 
    exactly $2$ colorful edges  is at most 5. 
         Let $m$ denote the minimal possible utility at equilibrium.
    
        Case 1: $m=0$. Let $x$ be the type which has an agent with utility 0. Then By Lemma \ref{lem:U_sharp_basic_property}(a), all other types have utility 3, so $\sw(L, U_\#) \ge 3(t-1)k = 3tk - 3k$.
    
        Case 2: $m=1$. Let $x$ be a type which has an agent with utility 1. By Lemma \ref{lem:U_sharp_basic_property}(a), all other types have utility at least 2. Also by Corollary \ref{cor:U_sharp_num_of_types}, there are at most 5 types whose agents can have utility exactly 2. For all remaining types, all agents have utility 3. Therefore 
            $\sw(L, U_\#)\ge k + 2 (5 k) + 3(t-6))k =  3tk - 7k$.
    
        Case 3: $m=2$. By Corollary \ref{cor:U_sharp_num_of_types}(a), there are at most 5 types whose agents can have utility exactly 2. For all remaining types, all agents have utility 3. Therefore            $\sw(L, U_\#) \ge 2 (5 k) + 3(t-5)k =  3tk - 5k$.
    
        Case 4: $m=3$. In this case $\sw(L, U_\#) \ge 3tk$.

        In all cases, we have  $\sw(L, U_\#) \ge (3t-7)k$, which  implies the claimed upper bound.   \qedhere
     \end{enumerate}
\end{proof}

\begin{theorem}\label{thm:4reg-U_sharp}
    Consider the swap game under $U_\#$ on $T_n$ with a set of $tk$ equitable agents belonging to $t$ types.
    \begin{enumerate}[(a)]
        \item For $t=2$, we have $\poa(\sw,U_\#, T_n) \eqas 5/3$;

        \item  For $t\ge 3$, we have
         $t/(t-1) \leqas  \poa(\sw,U_\#, T_n)  \leq \min\{ t/(t-13/4), \frac{4}{3} t/(t-1) \}$.
    \end{enumerate}
\end{theorem}

\begin{proof}
    \begin{enumerate}[(a)]
        \item In Assignment $L_{opt}^T$, every agent has utility 4, and $\sw(L_{opt}^T, U_\#) = 8k$. Consider now Assignment $L_\#^T$ in which the type sequence $2112212212$ is repeated in odd rows, while in the even rows, the type sequence $2112112211$ is repeated; in each row, the sequence is shifted to the right by one compared to the previous row. It is easy to verify that $L_\#^T$ is an equilibrium, and that in each sequence of length 10 described above, there are six agents with utility 2 and four with utility 3 which implies that $\sw(L_\#, U_\#) = 24 \times \frac{2k}{10} = 4.8k$. Thus $\poa(\sw,U_\#, T_n) \geq \frac{8k}{4.8k} = 5/3$.

        Next we show that in any equilibrium $L$, the social welfare is at least $4.8k$ which will show that the bound on PoA is tight.  Let $m$ denote the minimal utility among all agents.  Lemma~\ref{lem:U_sharp_basic_property}(e) implies $m\ge 1$.

        Case 1: $m=1$. If some type 1 agent has utility $1$, 
        Lemma \ref{lem:U_sharp_basic_property}(a)   implies that all agents of type 2  have utility at least $3$, hence, 
        $\ce(L) \ge 3k$ which implies $\sw(L, U_\#) = 2\ce \ge 6k$.

        Case 2: $m=2$. Recall that  $k_{ij}$  denotes the number of agents of type $i$ having utility exactly $j$. Suppose some type 1 agent $A$ has  utility exactly $2$, equivalently, there are exactly two colorful edges incident to $A$, hence in total there are $2k_{12}$ such edges Consider now the type 2 agents in such edges. Lemma \ref{lem:U_sharp_basic_property}(b) implies that such type 2 agents all have utility at least $3$. We now partition all these $2k_{12}$ colorful edges into two subsets: the ones where the type 2 agent has utility $3$, and the ones where the type 2 agent has utility $4$, and let $x_2, y_2$  denote the number of edges in each of these two subsets. Hence,
        \begin{equation}    \label{eq:U_sharp_LP1}
            x_2 + y_2 = 2k_{12}. 
        \end{equation}
        Since each type 2 agent of utility $3$ is `shared' by exactly three type 1 agents, we have,
        \begin{equation}    \label{eq:U_sharp_LP2}
            k_{23} \ge x_2 / 3. 
        \end{equation}
        Similarly, 
        \begin{equation}    \label{eq:U_sharp_LP3}
            k_{24} \ge y_2 / 4. 
        \end{equation}
        By a symmetrical argument, we have
        \begin{equation}    \label{eq:U_sharp_LP4}
            \begin{cases}
                x_1 + y_1 = 2 k_{22}, \\
                k_{13} \ge x_1 / 3;  \\ 
                k_{14} \ge y_1 / 4,
            \end{cases}
        \end{equation}
        where $x_1, y_1$ are defined analogously. Also, since $m=2$, we have $k_{11} = k_{21} = 0$, therefore
        \begin{equation}    \label{eq:U_sharp_LP5}
            \begin{cases}
                k_{12} + k_{13} + k_{14} = k; \\
                k_{22} + k_{23} + k_{24} = k.
            \end{cases}
        \end{equation}
        Now observe that the social welfare is 
            $2(k_{12} + k_{22}) + 3(k_{13} + k_{23})  + 4(k_{14} + k_{24})$
        Minimizing the above with the linear constraints \eqref{eq:U_sharp_LP1}-\eqref{eq:U_sharp_LP5} using an LP-solver, we obtain $\sw(L, U_\#) \ge 4.8 k$. 

        Case 3: $m\ge 3$. This automatically implies $\sw(L, U_\#) \ge 2k \cdot 3 = 6k$.

        In all cases, we have verified that $\sw(L, U_\#) \geq 4.8 k$ which completes the proof that $\poa(\sw,U_\#, T_n) = 5/3$.

        \item   Assignment $A_{opt}^T$ from Table~\ref{table:opt} has social welfare $tk$. On the other hand $A_{\tau}^T$ from Table~\ref{table:U-tau} is an equilibrium in which every agent of type other than $t$ has utility 4, and all but $o(k)$ agents of type $t$ are segregated. Therefore $\sw(A_{\#}^T, U_\#) = (t-1)k + o(k)$, hence $\poa({\sw},{U_\#}, T_n) \geq 3/2$. Also all but $o(k)$ agents have agents of at most one type in their neighborhood, so $\wdoit{(2)} = 0$.
        
        Let $L$ be an arbitrary equilibrium assignment. We will now show that  $\sw(L, U_\#) \ge \max\{4tk - 13k, 3tk - 3k\}$ which will prove the claimed upper bound on $\poa(\sw,U_\#, T_n)$.

    By Corollary~\ref{cor:U_sharp_num_of_types}, the number of types whose agents can have 
    exactly $2$ colorful edges  is at most 5.

    Let $m$ denote the minimum utility of an agent in $L$, and let $\alpha$ be a type such that at least one of the agents of type $\alpha$ has utility $m$. 

    Case 1: $m=0$. By Lemma \ref{lem:U_sharp_basic_property}, all  agents of all types other than $\alpha $ have utility 4. Therefore, 
        $\sw(L, U_\#) \ge 4(t-1)k = 4tk - 4k$.

    Case 2: $m=1$. By Lemma \ref{lem:U_sharp_basic_property}, all agents of types other than $\alpha$ have utility at least 3. By Corollary \ref{cor:U_sharp_num_of_types}, there are at most 7 types whose agents can have utility 3, and at most 
    \[
        \sw(L, U_\#) \ge k +  3 (7k) + 4(t-7)k
        =  4tk - 6k
    \]
    If we simply apply Lemma \ref{lem:U_sharp_basic_property}, we get
        $\sw(L, U_\#) \ge k + 3 (t-1)k = 3tk - 2k$.
    Hence,
        $\sw(L, U_\#) \ge \max\{4tk - 6k, 3tk - 2k\}$.

    Case 3: $m\ge 2$. By Lemma \ref{lem:U_sharp_basic_property}, all agents then have utility at least 2,  and Corollary \ref{cor:U_sharp_num_of_types}, agents of at most 3 types can have utility 2, and agents of at most 7 types can have utility 3. Therefore, 
    \[
        \sw(L, U_\#) \ge 2 (3k)   + 3 (7k)  k + 4 (t - 10) k
        =  4tk - 13k.
     \]
     We can also choose to apply Corollary \ref{cor:U_sharp_num_of_types} only with $h(2)$, then together with Lemma \ref{lem:U_sharp_basic_property}  we have
     \[
        \sw(L, U_\#) \ge 2 (3 k) + 3 \cdot (t- 3) k
        =  3tk - 3k.
     \]
     Hence,
        $\sw(L, U_\#) \ge \max\{4tk - 13k, 3tk - 3k\}$.

    To sum up, in all cases, we have  $\sw(L, U_\#) \ge \max\{4tk - 13k, 3tk - 3k\}$ which implies the claimed upper bound on $\poa(\sw,U_\#, T_n)$.      \qedhere
    \end{enumerate}
\end{proof}


\section{Experiments} \label{sec:experiments}

\begin{figure*}[ht]
        \centering
      \includegraphics[scale=.42]{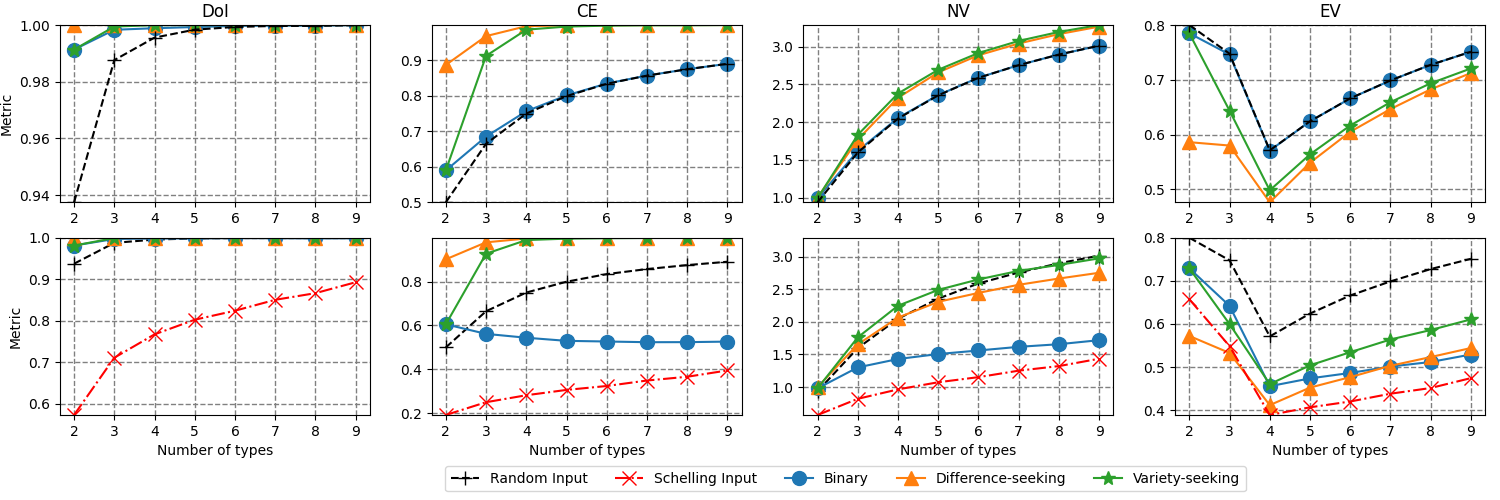}
        \caption{The first row is for random input, the second row for Schelling input. From left to right, the average $\doi, \ce, \nv, \ev$ (normalized) values of the equilibrium assignment reached by the swap games under $U_b$, $U_\#$, and $U_\tau$ are plotted. The values of the corresponding diversity measures present already in the inputs are also plotted. } 
        \label{fig:experiment}
    \end{figure*}

For every $U\in \{U_b, U_\#, U_\tau\}$, and for the number of types $t =2, \ldots, 9$, we ran the swap game on a 4-regular torus with two different initializations: the {\em random input}  and the {\em Schelling input}. The random input assigns to each vertex an agent of a type chosen uniformly at random  from $\{1,\ldots, t\}$.
The Schelling input is obtained by running a swap Schelling game using the similarity-seeking utility function of  \cite{schelling-journal} on the random input.
Each swap is chosen as follows: agents are ordered by their utilities, the agent with the smallest utility is picked, and we search through this ordered list of agents for a swap. Each experiment stops when there is no possible swap, i.e., it reaches an equilibrium. Finally, the result is evaluated with the diversity measures $\doi, \ce, \nv$, and $\ev$. In Figure \ref{fig:experiment}, $\ce$ is normalized as $\ce(L) / |E|$, and $\ev$ is normalized as $\ev(L)/\OPT$ where $\OPT$ corresponds to an assignment in which the $t$ types are evenly distributed in every neighborhood. For each type of input, we ran 50 simulations on 900 vertices and report the average result. In all plots, the higher the value, the more diversity. It can be seen in Figure 5 that the random input already generally exhibits high levels of diversity, while the Schelling input exhibits low levels of diversity, in both cases, the diversity (except for $\ev$) increases with the number of types.

The main findings are:  (1) Segregation is effectively removed when agents are diversity-seeking; however, strong diversity such as measured by evenness is still hard to achieve via $U_b, U_\#$ or $U_\tau$. \\
(2) Regardless of the starting input, the swap game under $U_\#$ performs  better than the game under $U_\tau$ for $\doi$ and $\ce$, while $U_\tau$ is better for $\nv$ and $\ev$. 

The second row of Figure \ref{fig:experiment} shows the results for Schelling inputs. We observe from the $\doi$ figure that at equilibrium of all the swap games, almost no agent is segregated. The $\ce$ figure shows that for agents at equilibrium under either $U_\#$ or $U_\tau$, every agent's neighborhood is almost $100\%$  colorful when $t\ge 4$ (the degree of the torus). The $\nv$ figure shows that $U_\#$ and $U_\tau$ achieve neighbourhood variety that initially surpasses that of the random assignment, though it appears to stabilize as $t$ increases. On the other hand, the $\ev$ figure shows that all three swap games do {\em worse} than the random assignment, e.g., $U_\tau$ which is the best among the three, lacks about $8 \sim 15\%$ evenness depending on the number of types, when compared with the random assignment. Also evenness is the most difficult to achieve when the number of types coincides with the degree of the graph.

Finally, we discuss two interesting phenomena. 
(1) All curves match the shape of the curve of the input, except the $\ce$ curve of $U_b$ for Schelling input which decreases slightly before appearing to stabilize at a little over 50\%. This could be due to the following factors: each swap under $U_b$ increases the value of $\ce$ by 8;  the number of swaps reduces as $t$ increases; $\ce$ in the Schelling input also increases gradually as $t$ increases. 
(2) For $U_\tau$ on  the Schelling input, the $\nv$ figure and $\ev$ figure show that an agent's neighborhood contains more types than that of a random assignment, but are distributed less evenly. This may be due to the fact that  evenness requires the even distribution of \emph{all} types, including the agent's type, while the $\ce$ figure shows that almost every agent's neighborhood is colorful, i.e., does not contain its own type.

Next we present some results on the more refined degree of integration measures. Recall that $\doic{(L,j)}$ is the fraction of agents with $j$ colorful edges incident on them and $\doit{(L,j)}$ is the fraction of agents with neighbors of $j$ types different from its own. We plot the average values of these measures over all considered assignments as $\doic{j}$ and $\doit{j}$. 

\begin{figure*}[ht]
        \centering
      \includegraphics[scale=.42]{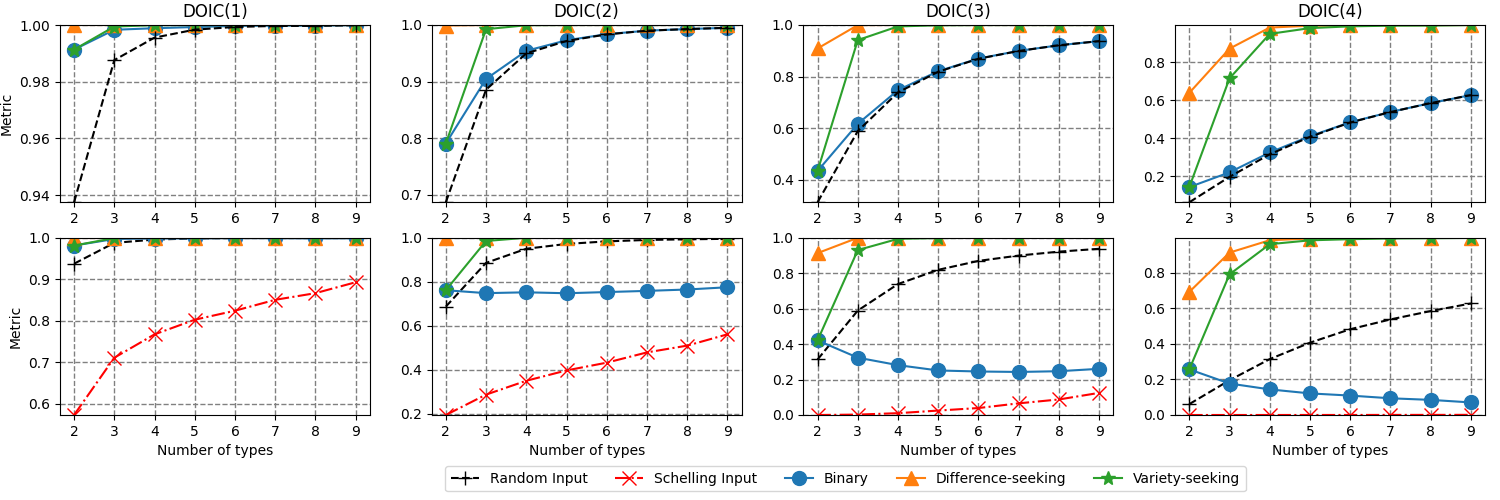}
        \caption{The first row is for random inputs, the second row is for Schelling inputs. From left to right, the average fraction of agents with 1, 2, 3, and 4 colorful edges incident on them at equilibrium under $U_b$, $U_\#$, and $U_\tau$, as well as the value in the input are plotted. } 
        \label{fig:experiment_DOICj}
    \end{figure*}

For the $\doic{(j)}$  measure, as shown in Figure~\ref{fig:experiment_DOICj},  we note the following: (1) On both random and Schelling inputs, the swap games under both $U_\#$ and $U_\tau$ perform better than $U_b$ on achieving a higher level of diversity (2) The type of input does not affect the performance much for $U_\tau$ and $U_\#$. However, while $U_b$ for a random input has roughly the same diversity as the initial input, on a Schelling input for higher values of $j$, it achieves much worse diversity than a random input.  (3) The performance of $U_\#$ is better than that of $U_\tau$, especially for smaller number of types. (4) Starting with Schelling inputs, both $U_\#$ and $U_\tau$ do better than a random assignment, especially for higher $j$.

\begin{figure*}[ht]
        \centering
      \includegraphics[scale=.42]{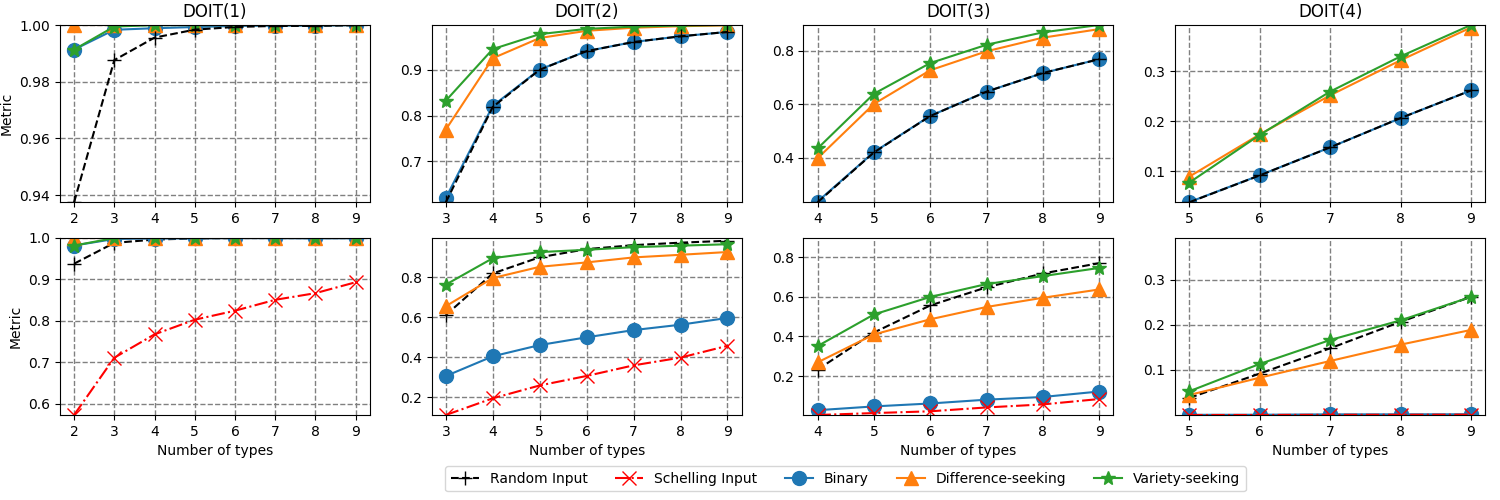}
        \caption{The first row is for random inputs, the second row is for Schelling inputs. From left to right, the average fraction of agents with 1, 2, 3, and 4 types of agents in their neighborhood at equilibrium under $U_b$, $U_\#$, and $U_\tau$, as well as the value in the input are plotted. }         \label{fig:experiment_DOITj}
    \end{figure*}

For the $\doit{(j)}$  measure, as shown in Figure~\ref{fig:experiment_DOITj}, we note that: (1) Both $U_\#$ and $U_\tau$ do better than $U_b$ on both types of inputs  (2)$U_\#$ and $U_\tau$ have similar performance on random inputs, and are not able to achieve high levels of diversity. For example, despite agents aiming to increase the number of types in their neighborhood, only between 10\% to 30\% of agents have 4 types in their neighborhood after running the swap game under $U_\tau$. (3) For Schelling inputs, $U_\tau$  has higher $\doit{j}$ for all values of $j>2$ than $U_\#$ (4) For higher values of $j$, the value of $\doit{j}$ after running the swap game under $U_\tau$ and $U_\#$ is very close to the initial value on a random input; this is in contrast with situation for $\doic{j}$, where values much better than for random inputs are achieved.


\section{Conclusions and Discussion} \label{sec:conclusions}

We proposed three new diversity-seeking swap games and several diversity measures and proved  bounds on the price of anarchy and stability for our diversity measures in these games. Our theoretical and simulation results indicate that while segregation can be effectively removed if agents are diversity-seeking, strong diversity such as measured by evenness might still be hard to achieve if agents' utilities are given by $U_b, U_\#$ or $U_\tau$. 

Many of our bounds are tight, but it remains to show tight bounds on the PoA for colorful edges for swap games under $U_\#$ and $U_\tau$. Studying other utility functions or diversity measures is an interesting avenue of research, including some mix of $U_\#$ and $U_\tau$. It would also be interesting to study other swap rules, such as the perturbed Schelling model, or bounded rational agents \cite{zhang2004integrationist,young2020individual}).

\bibliographystyle{plain}
\bibliography{mybib}

\end{document}